\documentclass[prd,superscriptaddress,twocolumn,nofootinbib,showpacs,preprintnumbers,floatfix]{revtex4}

\usepackage{mathrsfs}
\usepackage{amsfonts,amssymb,amsmath}
\usepackage{graphicx,epsfig}
\usepackage{multirow}

\def\fsu5{$\cal{F}$-$SU(5)$}

\def\m1half{$M_{1/2}$}
\def\m3half{$M_{3/2}$}
\def\m32{$M_{32}$}

\def\htmiss{$H_{\rm T}^{\rm miss}$~}

\begin{document}

\title{Has SUSY Gone Undetected in 9-jet Events? \\ A Ten-Fold Enhancement in the LHC Signal Efficiency}

\author{Tianjun Li}

\affiliation{Key Laboratory of Frontiers in Theoretical Physics, Institute of Theoretical Physics,
Chinese Academy of Sciences, Beijing 100190, P. R. China }

\affiliation{George P. and Cynthia W. Mitchell Institute for Fundamental Physics and Astronomy,
Texas A$\&$M University, College Station, TX 77843, USA }

\author{James A. Maxin}

\affiliation{George P. and Cynthia W. Mitchell Institute for Fundamental Physics and Astronomy,
Texas A$\&$M University, College Station, TX 77843, USA }

\author{Dimitri V. Nanopoulos}

\affiliation{George P. and Cynthia W. Mitchell Institute for Fundamental Physics and Astronomy,
Texas A$\&$M University, College Station, TX 77843, USA }

\affiliation{Astroparticle Physics Group, Houston Advanced Research Center (HARC),
Mitchell Campus, Woodlands, TX 77381, USA}

\affiliation{Academy of Athens, Division of Natural Sciences,
28 Panepistimiou Avenue, Athens 10679, Greece }

\author{Joel W. Walker}

\affiliation{Department of Physics, Sam Houston State University,
Huntsville, TX 77341, USA }


\begin{abstract}

On the heels of the first analysis of LHC data eclipsing the inverse femtobarn integrated luminosity
milestone, we undertake a detailed comparison of the most recent experimental results with Monte Carlo simulation
of the full ``bare-minimally constrained'' parameter space of the class of supersymmetric models which go by the name of
No-Scale \fsu5.  We establish the first sparticle exclusion boundaries on these models, finding that the LSP mass should
be at least about $92$~GeV, with a corresponding boundary gaugino mass $M_{1/2}$ above about $485$~GeV. In contrast to the higher mass constraints established for the CMSSM, we find the minimum exclusion boundary on the \fsu5 gluino and heavy squark masses resides in the range of 658-674 GeV and 854-1088 GeV, respectively, with a minimum light stop squark mass of about 520 GeV.  Moreover, we show that elements of the surviving parameter space not only escape the onslaught of LHC data which is currently decimating the standard mSUGRA/CMSSM benchmarks, but are further able to \textit{efficiently explain} certain tantalizing production excesses over the SM background which have been reported by the CMS collaboration.  We also extend this study comparatively to five distinct collider energies and four specific cut methodologies, including a proposed set of
selection cuts designed to reveal the natural ultra-high jet multiplicity signal associated with the stable mass
hierarchy $m_{\tilde{t}} < m_{\tilde{g}} < m_{\tilde{q}}$ of the \fsu5 models.  By so doing, we demonstrate that a
rather stable enhancement in model visibility, conservatively of order ten, may be attained by adoption of these cuts,
which is sufficient for an immediate and definitive testing of a majority of the model space using only the existing LHC
data set.  We stress the point that habits established in lower jet multiplicity searches do not necessarily carry over
into the ultra-high jet multiplicity search regime.

\end{abstract}

\pacs{11.10.Kk, 11.25.Mj, 11.25.-w, 12.60.Jv}

\preprint{ACT-12-11, MIFPA-11-38}

\maketitle


\section{Introduction and Motivation}

The task given to experimentalists is one of exceeding difficulty, and the
position of the goalposts is in constant flux.  Yesterday's sensation is tomorrow's
calibration, as the maxim goes, and signals not long ago heralded simply for
the seeing, come soon to be reclassified as background, and must become
sufficiently well resolved that they may be seen {\it past}.  Seeking out the tails
of statistical distribution tails, the right tool is essential.  If one wishes to
find needles in a haystack, a magnet may prove more useful than a pitchfork.
No one much doubts that the LHC is a suitable tool, or that the forthcoming energy
doubling to $\sqrt{s} = 14$~TeV will make it ever much more so, and indeed many times
more so than twice.  However, there are certain complementary tools in the experimental
arsenal which may be deployed for a cost somewhat less than US \$10 billion.

The most significant such tool is the data selection cut.  Simply speaking, one
must isolate, or select, potential outcomes which are accessible to the desired
signal, but inaccessible to, or at least substantially unlikely for, the competing
background.  It may actually be beneficial to discard even a large quantity of signal
events from regions of phase space which are background dominated, in favor of a smaller
quantity of retained events of an unusual character which may be uniquely differentiated.
A second layer of selection filtering may typically be applied to eliminate faked signals
attributable to the persistent incompleteness and occasional fallibility of detector
measurements.  Careful tailoring of the cuts to the sought signal may readily account for
orders of magnitude of relative signal enhancement, and great effort is thus expended in
this pursuit, commensurate with the weight of potential benefit that the cuts employed
may leverage against the great cost and effort of the project at large.

Cuts highly specialized for a single given task are though not guaranteed
to be suitable for any secondary purpose, and worse, may in fact mask or obscure
the detection of alternatively feasible physical models.  The cuts featured most
prominently in the early LHC studies are geared toward evocation of minimal supergravity (mSUGRA)
and the Constrained Minimal Supersymmetric Standard Model (CMSSM).  This is neither surprising nor
unreasonable; chasing plurality without necessity is a fool's errand.  However, the
moment of escalation may be at hand with the recent announcement that a definitive
marker of supersymmetry (SUSY) at the LHC escapes the scrutiny of the first complete inverse femtobarn
of integrated luminosity~\cite{PAS-SUS-11-003}.  It was concluded by one particular study~\cite{Strumia:2011},
based on just the first $165~{\rm pb}^{-1}$ of data, that more than $99 \%$ of the
minimal parameter space had been already disfavored, and similarly dire pronouncements,
claiming an even greater scope, may be expected to follow the most recently reported results.
Most significantly, as exclusion boundaries on heavy squarks creep above the TeV level,
the very {\it raison d'\^etre} of SUSY itself becomes imperiled;  as a solution to the stabilization of the
gauge hierarchy of the electroweak scales, and as an appeal to naturalness against the spectre of fine tuning,
SUSY should embody sparticle mass splittings which are not greater than the range of several hundreds of GeV,
or at most about one TeV.

In the present work, we will perform an explicit comparison of the model space of a construction dubbed
No-Scale \fsu5 against the most recently reported experimental results, establishing the first sparticle mass
limits which apply to these models, and emphasizing that, by mimicking the published CMS cutting methodology,
we may further efficiently account within the remaining parameter space for the statistical
excesses (though insufficient for formal discovery) which have been observed above the Standard Model
(SM) expectation~\cite{PAS-SUS-11-003}.  We will then proceed by detailed steps toward the 
central conclusion of our current analysis, namely that the No-Scale \fsu5 model experiences on the order of
a ten-fold enhancement in visibility, and possibly even substantially more, by the application of cuts tuned to its
distinctive ultra-high jet signal; this enhancement is more than sufficient to claim immediate SUSY discovery within
large portions of the model, using only the existing data accumulation.

Whether this specific model should ultimately be shown to be correct or incorrect, it serves here nonetheless
as an immediate and practical warning against any misconception that limits currently being established by
the ATLAS and CMS collaborations within the mSUGRA/CMSSM context may be globally extrapolated onto
the underlying framework of SUSY itself.  A more comprehensive probe of the SUSY model space will
necessarily entail application of a wider variety of selection criteria against the wealth of accumulated
raw data.  After all, we remind the reader that CMS does indeed still stand for ``Compact Muon Solenoid'',
and most certainly not for ``Constrained Minimal Supersymmetry''.

\section{Foundations of No-Scale \fsu5}

Recently, we have studied in some substantial detail a model by the name of No-Scale
\fsu5~\cite{Li:2010ws,Li:2010mi,Li:2010uu,Li:2011dw,Li:2011hr,Maxin:2011hy,Li:2011xu,Li:2011in,Li:2011gh,Li:2011rp},
constructed upon the tri-podal foundation of the ${\cal F}$-lipped $SU(5)$
Grand Unified Theory (GUT)~\cite{Barr:1981qv,Derendinger:1983aj,Antoniadis:1987dx},
two pairs of hypothetical TeV scale vector-like SUSY multiplets with origins in
${\cal F}$-theory~\cite{Jiang:2006hf,Jiang:2009zza,Jiang:2009za,Li:2010dp,Li:2010rz},
and the dynamically established boundary conditions of No-Scale
Supergravity~\cite{Cremmer:1983bf,Ellis:1983sf, Ellis:1983ei, Ellis:1984bm, Lahanas:1986uc}.
For a more complete review, the reader is directed to the appendix of Ref.~\cite{Maxin:2011hy},
and to the references therein.

An accumulating body of work argues that the speculative components
of No-Scale \fsu5 offer actually a clear path toward reduced plurality in postulate, via
the singularly natural treatment of such disparate concerns as radiative electroweak
symmetry breaking (EWSB)~\cite{Ellis:1983bp} and stabilization of the gauge hierarchy, precision coupling 
unification~\cite{Ellis:1990zq, Ellis:1990wk, Amaldi:1991cn, Langacker:1991an, Anselmo:1991uu, Anselmo:1992jb}\footnote{Such precise gauge unification does not occur in non-supersymmetric $SU$(5)~\cite{Costa:1987qp}}
and reconciliation of the ``little hierarchy'' between the GUT and Planck
scales~\cite{Lopez:1995cs,Jiang:2006hf,Li:2010dp}, suppression of dimension five proton decay by colored
Higgsino exchange~\cite{Antoniadis:1987dx}, electroweak doublet-triplet splitting by the missing partner
mechanism~\cite{Antoniadis:1987dx}, a neutralino cold dark matter (CDM)
candidate~\cite{Ellis:1983wd, Ellis:1983ew, Goldberg:1983nd},
appropriately small seesaw neutrino masses~\cite{gellmann,yanagida,Georgi:1979dq},
chiral GUT higgs representations and gauge symmetry breaking by flux
activation~\cite{Beasley:2008dc,Beasley:2008kw,Font:2008id,Chen:2009me,Jiang:2009zza,Jiang:2009za,Li:2009cy},
realization of the gravitational decoupling scenario~\cite{Jiang:2009zza,Jiang:2009za},
narrow refinement of the electroweak (EW) Higgs vacuum expectation value (VEV) ratio $\tan \beta$,
the dynamic origin of a single SUSY breaking mass, cosmological flatness~\cite{Cremmer:1983bf}, and the
monopole problem, all within a framework granted the {\it imprimatur} of a string motivated origination.

The model demonstrates a high precision of predictive constraint, especially
in the relatively narrow parameterization freedom of the SUSY sparticle sector, as enforced by the No-Scale
boundary conditions, and in particular, the non-trivial vanishing of the Higgs bilinear soft term $B_\mu$ at
the high scale. Notably, this scenario appears to comes into its own only when applied at an elevated scale,
approaching the Planck mass~\cite{Ellis:2001kg,Ellis:2010jb}.  Likewise, $M_{\cal F}$, the point of the second stage flipped
$SU(5)\times U(1)_{\rm X}$ unification, emerges in turn as a suitable candidate scale only when substantially
decoupled from the primary GUT scale unification of $SU(3)_{\rm C} \times SU(2)_{\rm L}$ via the modification to the
renormalization group equations (RGEs) from the extra ${\cal F}$-theory vector multiplets~\cite{Li:2010ws,Li:2010mi}.
This interdependence highlights the mutually essential roles of the three foundational postulates, and the
manner by which they conspire to reduce rather than enlarge the level of uncertainty in the model's
predicted phenomenology.

We will here show that regions of the bare-minimally constrained~\cite{Li:2011xu} parameter
space of No-Scale \fsu5, simultaneously consistent with the measured top-quark mass $m_{\rm t}$, the No-Scale boundary conditions,
radiative EWSB, the centrally observed WMAP7 CDM relic density~\cite{Komatsu:2010fb},
and precision LEP constraints on the lightest CP-even Higgs boson $m_{h}$~\cite{Barate:2003sz,Yao:2006px} and
other light SUSY chargino and neutralino mass content, remain viable even after announcement of the upgraded LHC data sets.
The most favorable regions that include secondary bounds on the flavor changing neutral current $(b \rightarrow s\gamma)$ process and on contributions to the muon anomalous magnetic moment $(g-2)_\mu$ are automatically satisfied for at least regional intersections within this space,
while limits on the rare decay $B_s^0 \rightarrow \mu^+ \mu^-$~\cite{Chatrchyan:2011kr} are satisfied for the entire model space, all of which is moreover consistent with spin-independent~\cite{Aprile:2011hi} and spin-dependent~\cite{Tanaka:2011uf} scattering cross-section bounds on Weakly Interacting Massive Particles (WIMPs), and ongoing collider searches for a Higgs signal.
We attribute this remarkable survival chiefly to the characteristically stable sparticle mass hierarchy
$m_{\tilde{t}} < m_{\tilde{g}} < m_{\tilde{q}}$ of a light stop and gluino, both
comfortably lighter than all other squarks.  This hierarchy allows No-Scale \fsu5 to evade collider
limits on light squark masses much more nimbly than CMSSM constructions with comparably light Lightest
Supersymmetric Particles (LSPs).

Most critically, this convergence of theoretical efficiency, tight predictive constraint, and adroit evasion
of the crossfire of current experimental results, is further coupled, somewhat paradoxically, to an encompassing
air of imminent testability.  No-Scale \fsu5 has escaped the advancing LHC exclusions not by being vague or by making
predictions which are inaccessible to contemporary search capabilities, but rather by hiding in plain sight.
The full parameter space features a dominantly Bino LSP at a purity greater than 99.7\%,
which is automatically safe with respect to existing DM direct detection searches, as led by the XENON100~\cite{Aprile:2011hi}
collaboration, yet well positioned for a near-term shot at the discovery and classification of a CDM
candidate~\cite{Li:2011in}.  With regards to the LHC collider effort, the same properties of the sparticle mass
hierarchy which cause No-Scale \fsu5 to fare rather poorly under conventional search strategies simultaneously affords
a stably definitive collider signal of ultra-high multiplicity jet events~\cite{Maxin:2011hy,Li:2011hr} which 
yields a dramatic resonant enhancement in visibility under selection cuts tuned to its own peculiar character.
This latter point will be supported by the detailed demonstration that adoption of a non-standard ultra-high jet multiplicity selection
criteria can yield on the order of a ten times enhancement in the detection efficiency of No-Scale \fsu5, within the
existing physical constraints of the collider apparatus.

Like Michelangelo's David, ensconced still by raw marble before application of the hammer and chisel,
our prize may stand already before us; all we need yet do is cut away that which is not David.

\section{No-Scale \fsu5 in the Light of a $1.1~{\rm fb}^{-1}$ Integrated Luminosity\label{sct:1p1ifb}}

\begin{figure*}[htp]
        \centering
        \includegraphics[width=0.80\textwidth]{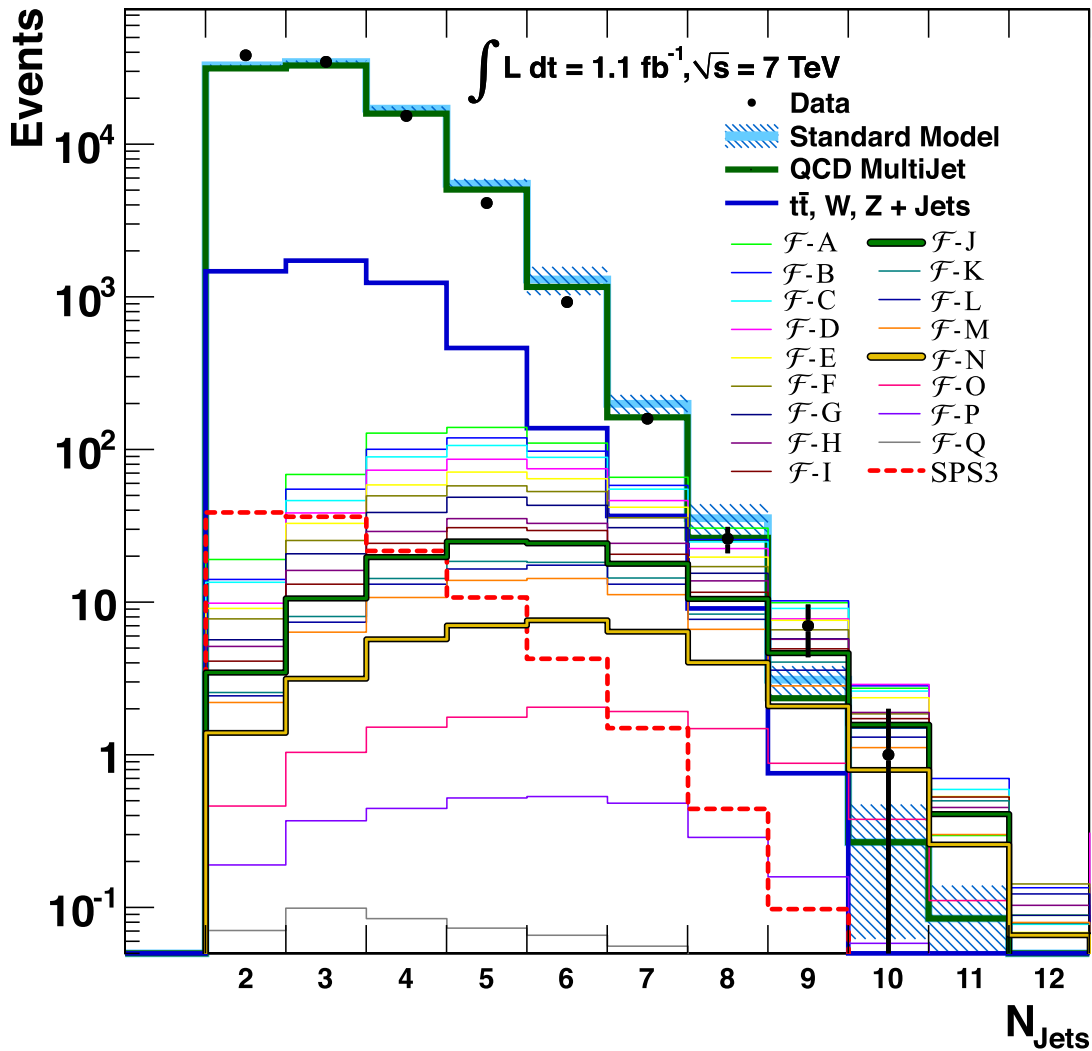}
        \caption{ The CMS Preliminary 2011 signal and background statistics for $1.1~{\rm fb}^{-1}$ of integrated luminosity
	at $\sqrt{s} = 7$~TeV, as presented in \cite{PAS-SUS-11-003}, are reprinted with an overlay consisting of a Monte Carlo
	collider-detector simulation of the No-Scale \fsu5 model space benchmarks of Table~(\ref{tab:compareN}), and the
	CMSSM benchmark SPS3. An experimentally favored region consistent with the bare-minimal experimental constraints of~\cite{Li:2011xu} and both the $(b \rightarrow s\gamma)$ process and contributions to the muon anomalous magnetic moment $(g-2)_\mu$ within the \fsu5 model space is represented by the emphasized gold contour. The emphasized green contour indicates the estimated lower bound on the parameter space after application of the CMS $1.1~{\rm fb}^{-1}$ analysis. The plot counts events per jet multiplicity, with no cut on $\alpha_{\rm T}$.
        } \label{fig:N_Jets}
\end{figure*}

\begin{figure*}[htp]
        \centering
        \includegraphics[width=0.45\textwidth]{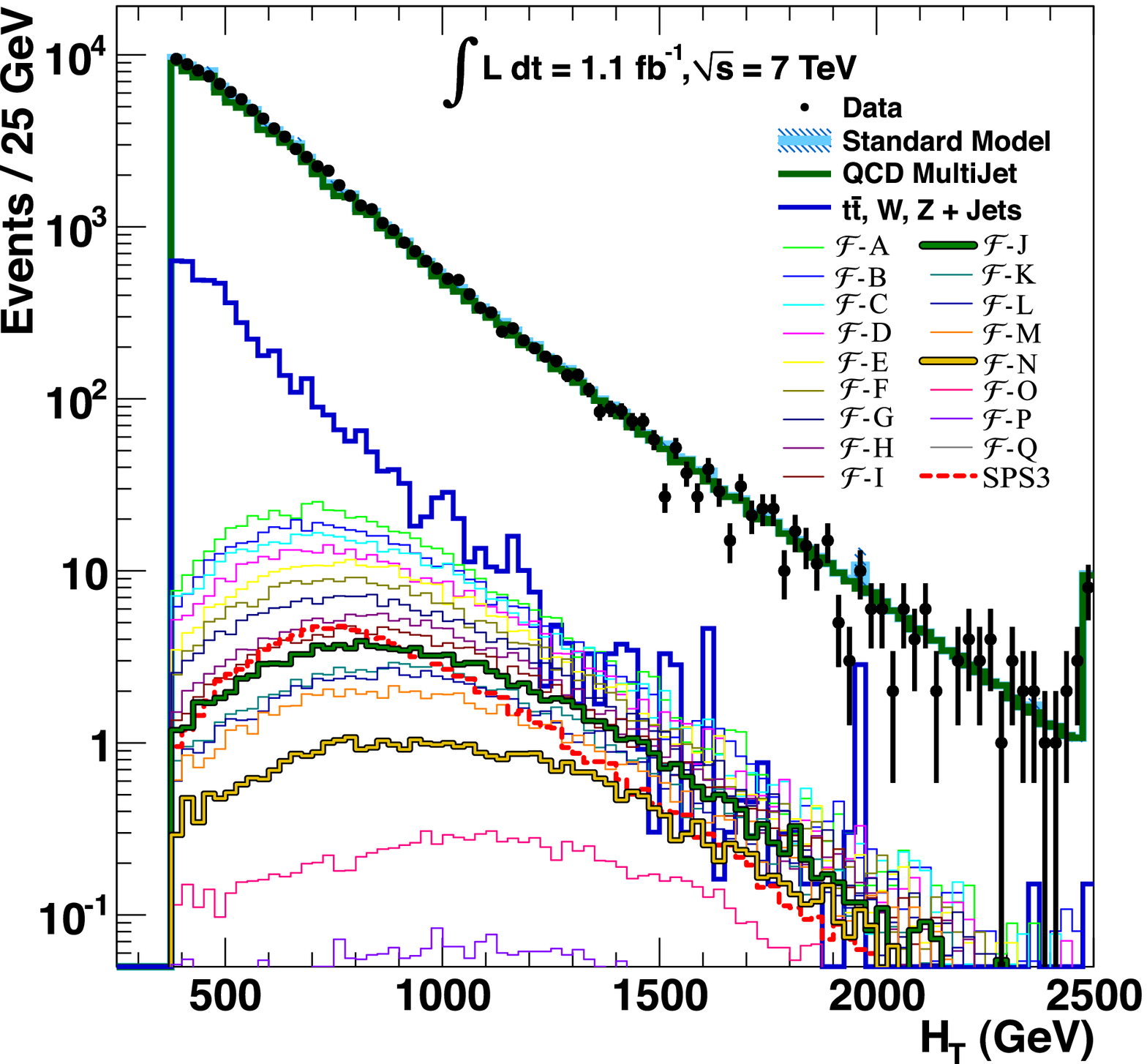}
	\hspace{0.05\textwidth}
        \includegraphics[width=0.45\textwidth]{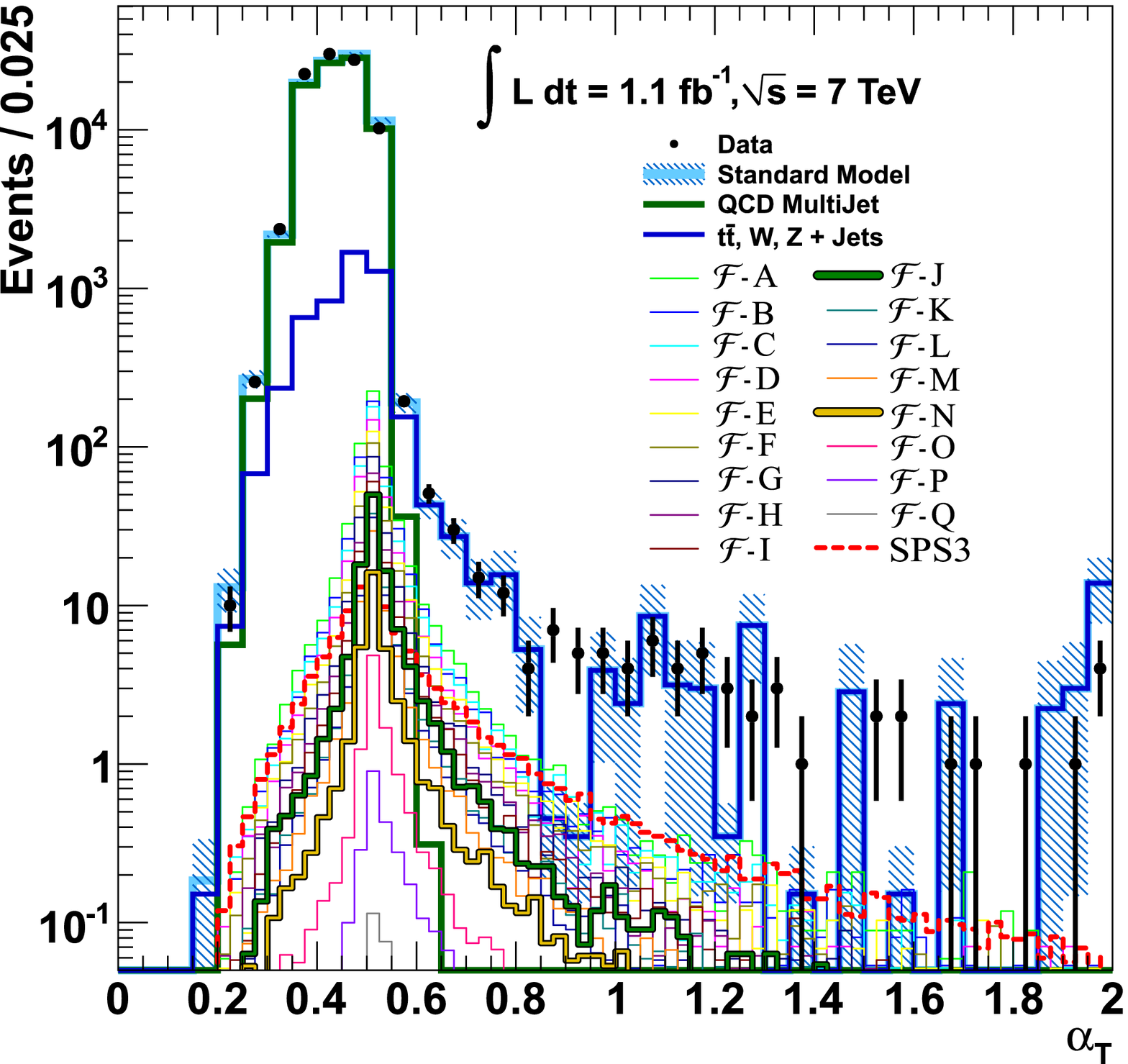} \\
	\vspace{0.02\textwidth}
        \includegraphics[width=0.42\textwidth]{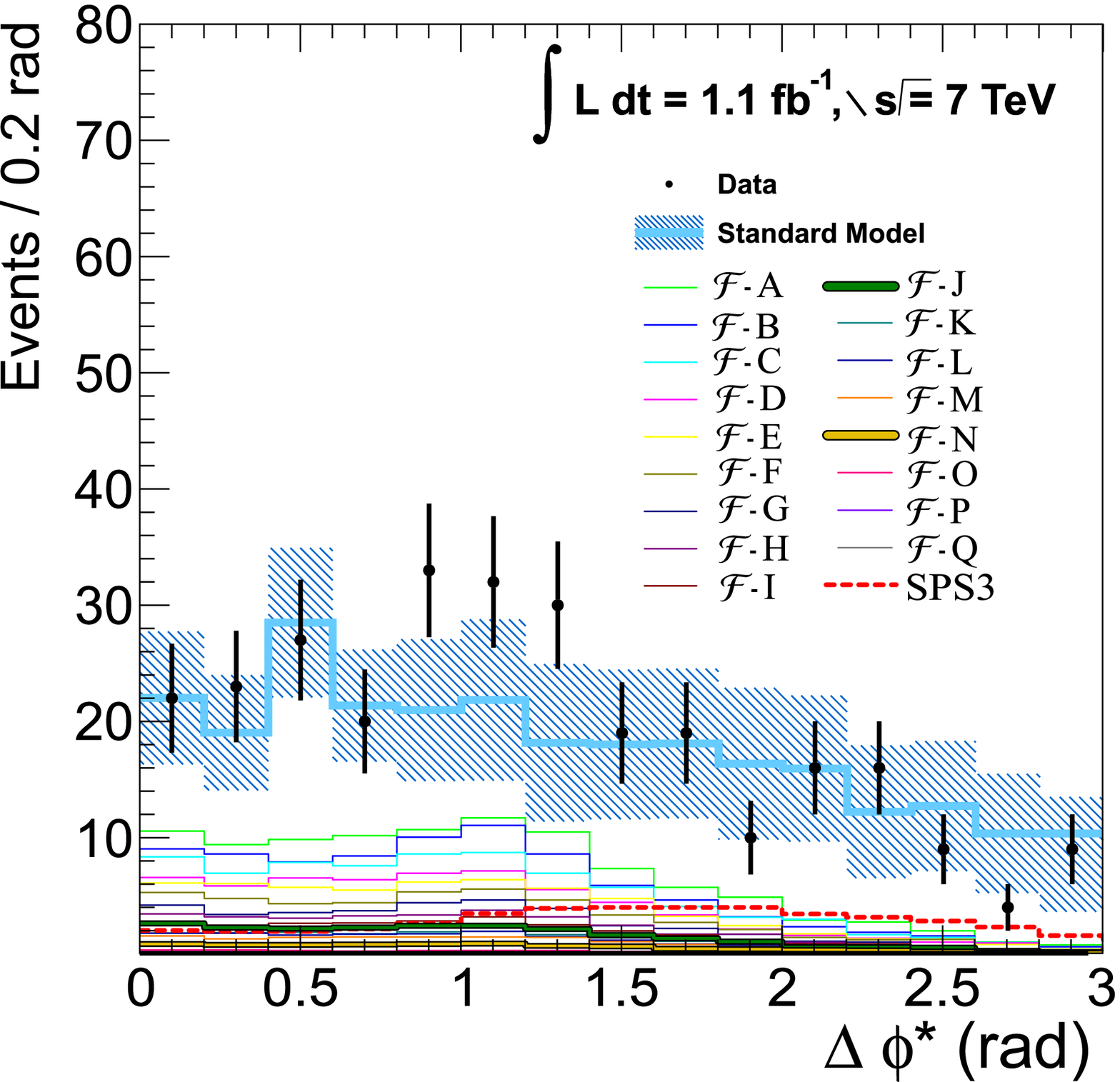}
	\hspace{0.05\textwidth}
        \includegraphics[width=0.42\textwidth]{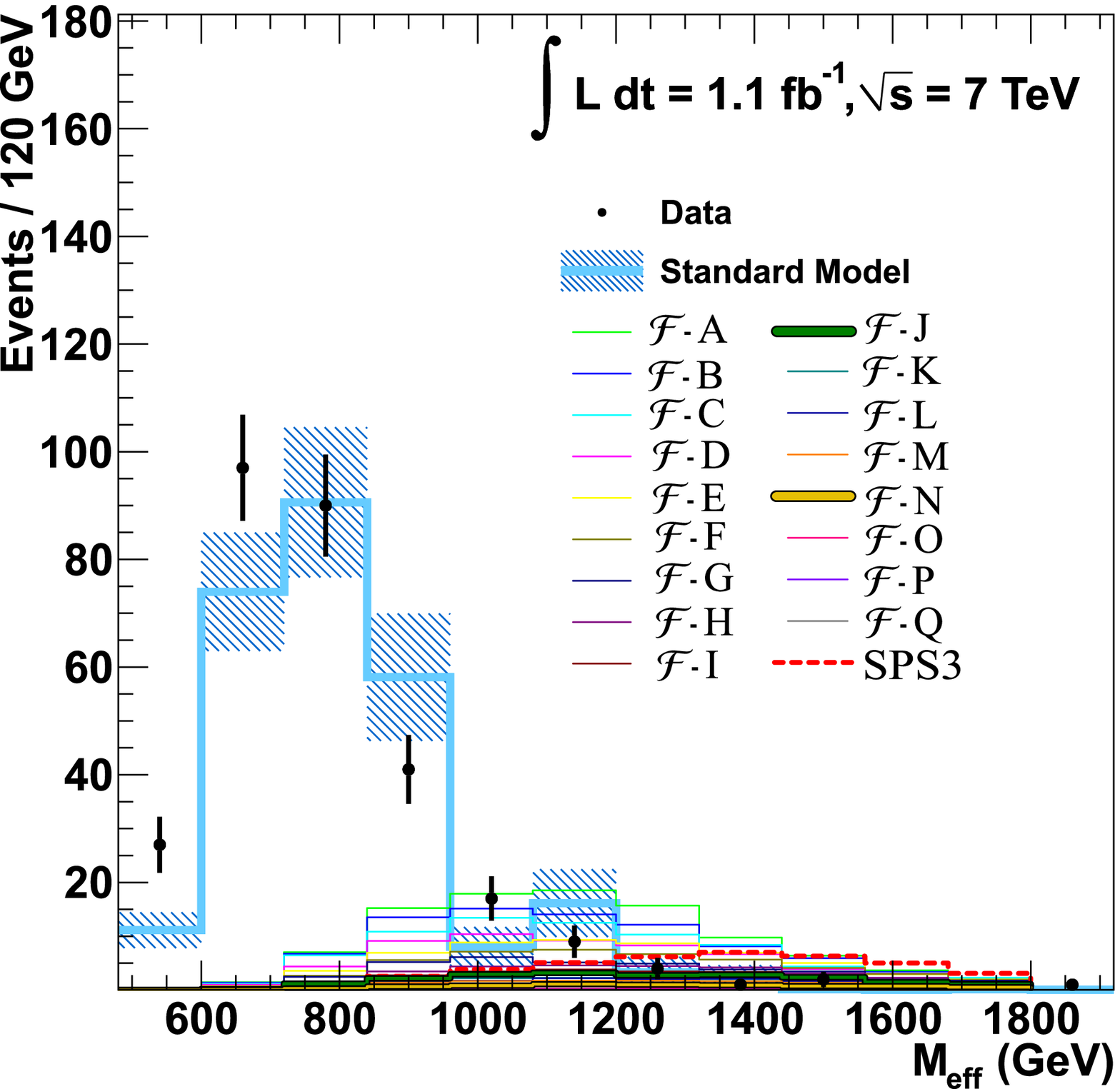}
        \caption{ The CMS Preliminary 2011 signal and background statistics for $1.1~{\rm fb}^{-1}$ of integrated luminosity
	at $\sqrt{s} = 7$~TeV, as presented in \cite{PAS-SUS-11-003}, are reprinted with an overlay consisting of a Monte Carlo
	collider-detector simulation of the No-Scale \fsu5 model space benchmarks of Table~(\ref{tab:compareN}), and the
	CMSSM benchmark SPS3.	An experimentally favored region consistent with the bare-minimal experimental constraints of~\cite{Li:2011xu} and both the $(b \rightarrow s\gamma)$ process and contributions to the muon anomalous magnetic moment $(g-2)_\mu$ within the \fsu5 model space is represented by the emphasized gold contour. The emphasized green contour indicates the estimated lower bound on the parameter space after application of the CMS $1.1~{\rm fb}^{-1}$ analysis. Histograms are displayed for the four statistics $H_{\rm T}, \alpha_{\rm T}, \Delta \phi^*, {\rm and } M_{\rm eff}$. The lower two plots feature a cut on $\alpha_{\rm T} \ge 0.55$, while this is suppressed in the upper two plots.} \label{fig:4plex}
\end{figure*}

\begin{figure*}[htp]
        \centering
        \includegraphics[width=0.80\textwidth]{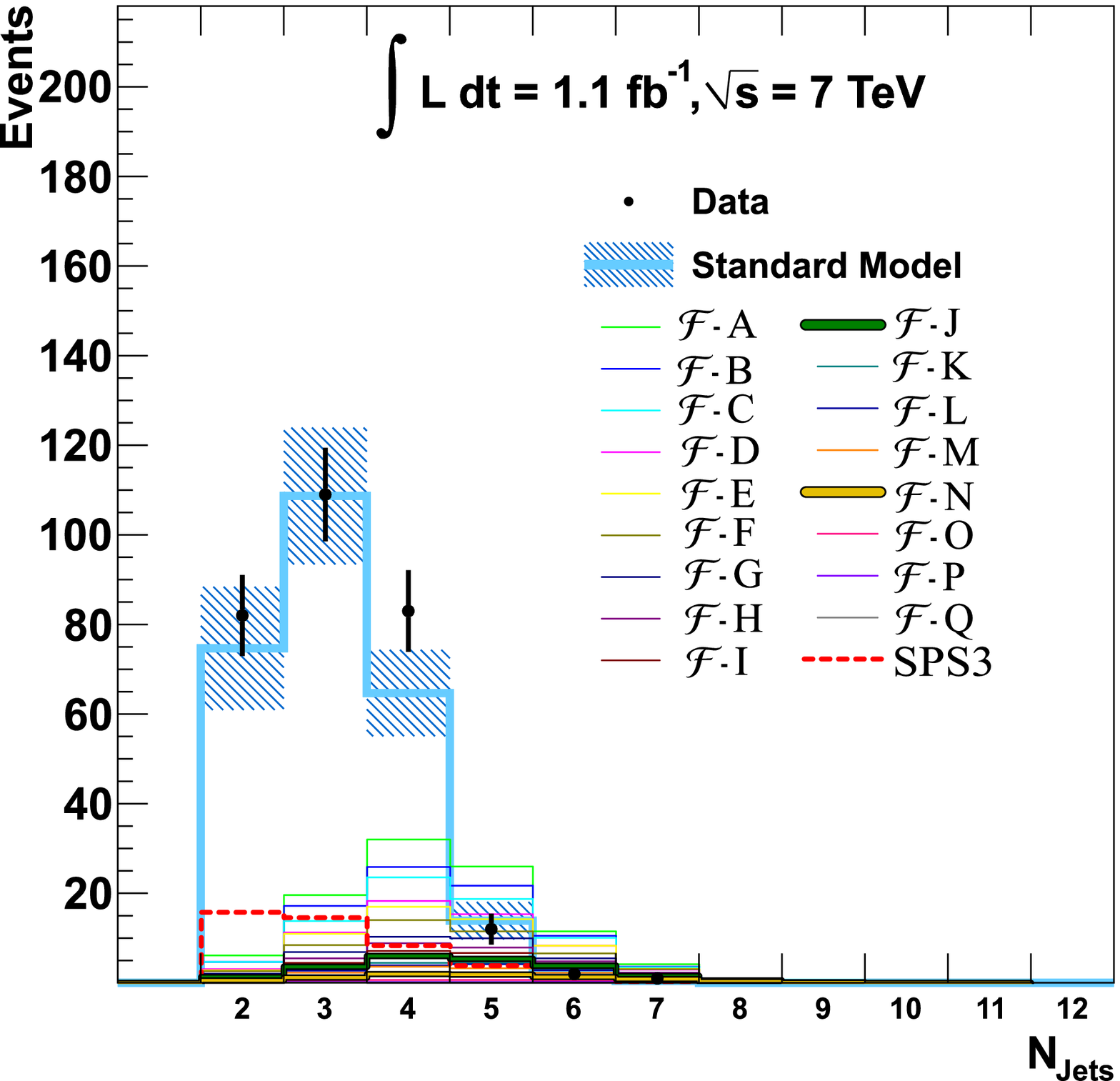}
        \caption{ The CMS Preliminary 2011 signal and background statistics for $1.1~{\rm fb}^{-1}$ of integrated luminosity
	at $\sqrt{s} = 7$~TeV, as presented in \cite{PAS-SUS-11-003}, are reprinted with an overlay consisting of a Monte Carlo
	collider-detector simulation of the No-Scale \fsu5 model space benchmarks of Table~(\ref{tab:compareN}), and the
	CMSSM benchmark SPS3.	An experimentally favored region consistent with the bare-minimal experimental constraints of~\cite{Li:2011xu} and both the $(b \rightarrow s\gamma)$ process and contributions to the muon anomalous magnetic moment $(g-2)_\mu$ within the \fsu5 model space is represented by the emphasized gold contour. The emphasized green contour indicates the estimated lower bound on the parameter space after application of the CMS $1.1~{\rm fb}^{-1}$ analysis. The plot counts events per jet multiplicity, with a cut on $\alpha_{\rm T} \ge 0.55$.} \label{fig:N_Jets_alpha_T}
\end{figure*}

\begin{figure*}[htp]
        \centering
        \includegraphics[width=0.80\textwidth]{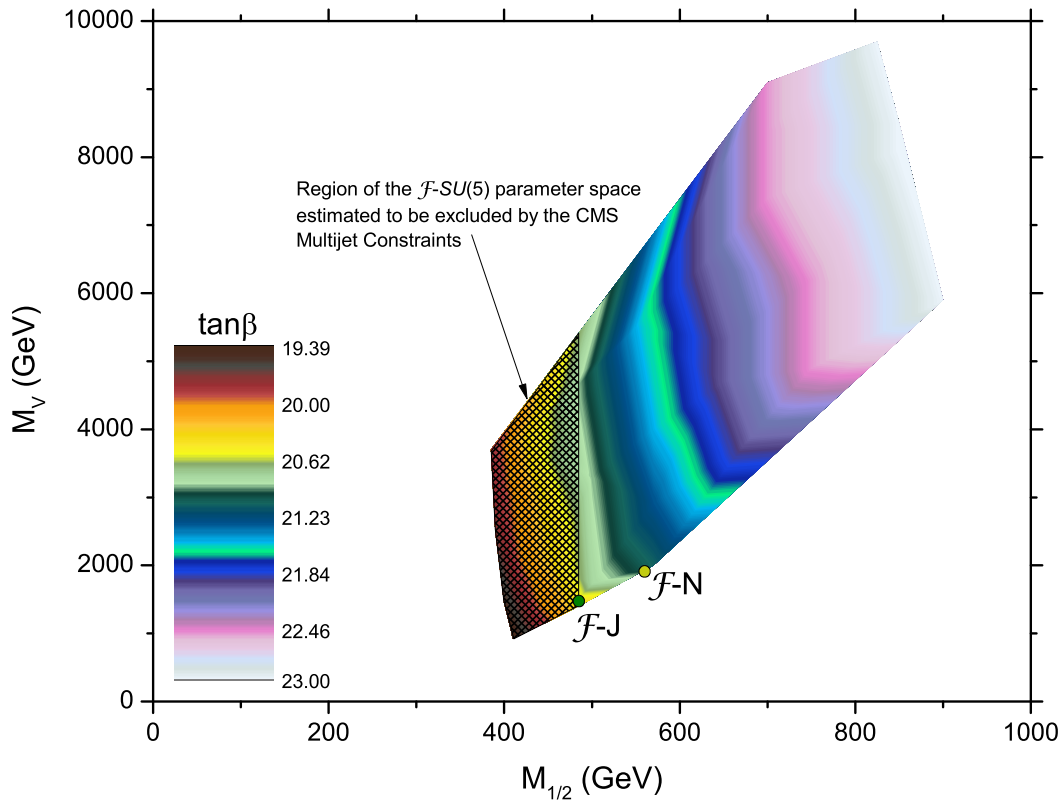}
        \caption{ The bare-minimally constrained parameter space of No-Scale \fsu5 is depicted as a function of
	the gaugino boundary mass $M_{1/2}$, the vector-like mass $M_{\rm V}$, and via the color key, the ratio
	of Higgs VEVs $\tan \beta$.  The region estimated to be disfavored by the first inverse femtobarn of integrated
	LHC luminosity has been marked with crosshatch. An experimentally favored region consistent with the bare-minimal experimental constraints of~\cite{Li:2011xu} and both the $(b \rightarrow s\gamma)$ process and contributions to the muon anomalous magnetic moment $(g-2)_\mu$ within the \fsu5 model space is represented by the gold point. The green point is a representative benchmark that indicates the estimated lower bound on the parameter space after application of the CMS $1.1~{\rm fb}^{-1}$ analysis.}
	\label{fig:Wedge}
\end{figure*}

\begin{figure*}[htp]
        \centering
        \includegraphics[width=0.80\textwidth]{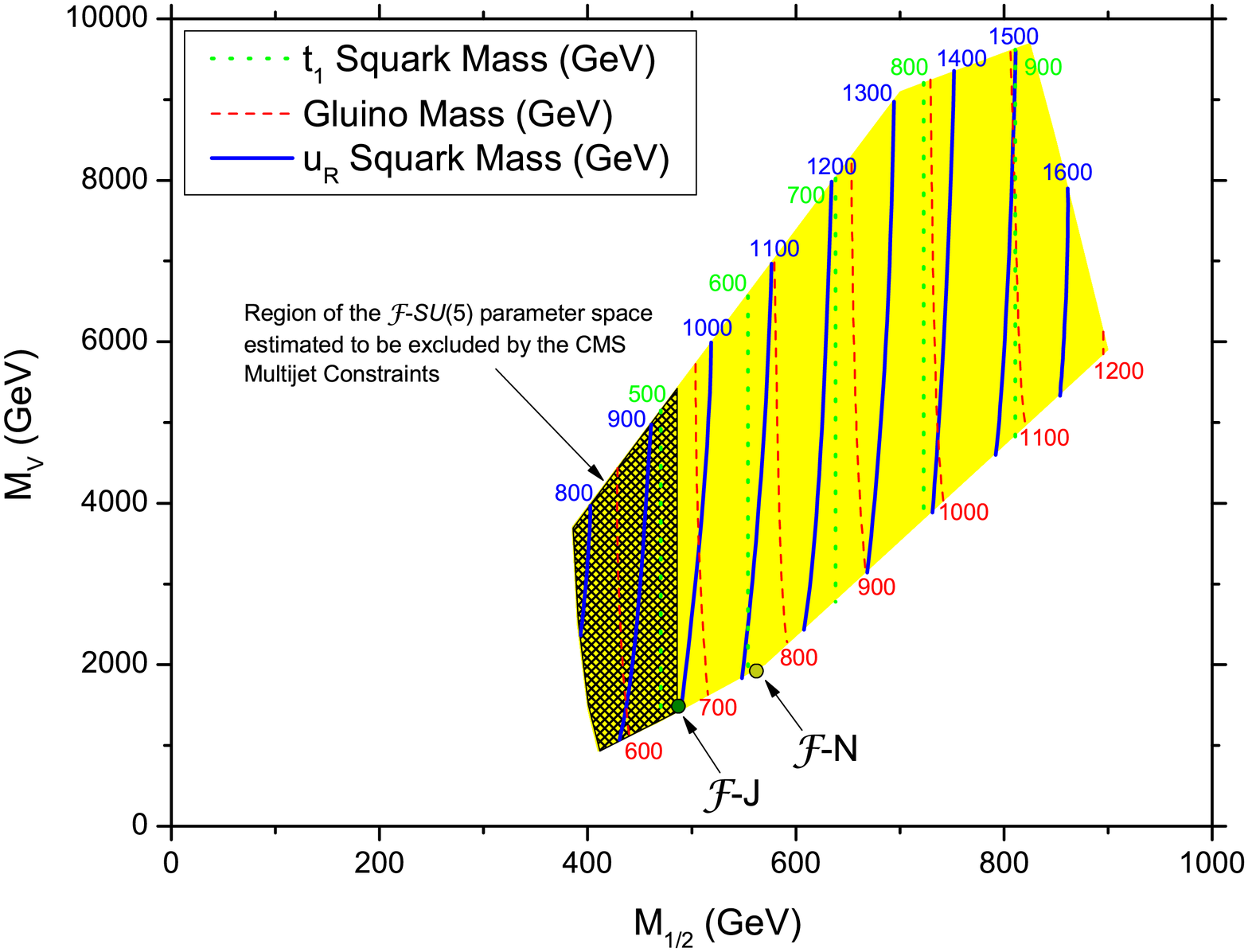}
        \caption{ The bare-minimally constrained parameter space of No-Scale \fsu5 is depicted as a function of
	the gaugino boundary mass $M_{1/2}$, the vector-like mass $M_{\rm V}$, and via the solid, dashed, and dotted contour lines, the masses in GeV of the light stop squark $\widetilde{t}_{1}$, gluino $\widetilde{g}$, and right-handed up squark $\widetilde{u}_{R}$. The upper bound on the region estimated to be disfavored by the first inverse femtobarn of integrated LHC luminosity has been marked with the crosshatched pattern. An experimentally favored region consistent with the bare-minimal experimental constraints of~\cite{Li:2011xu} and both the $(b \rightarrow s\gamma)$ process and contributions to the muon anomalous magnetic moment $(g-2)_\mu$ within the \fsu5 model space is represented by the gold point. The green point is a representative benchmark that indicates the estimated lower bound on the parameter space after application of the CMS $1.1~{\rm fb}^{-1}$ analysis. The lower bound on $m_{\widetilde{g}}$ for the entire model space occurs in the range of 658-674 GeV, while the lower bound on $m_{\widetilde{u}_{R}}$ transpires in the span of 956-1000 GeV. The lower bound for all the heavy squarks ($m_{\widetilde{t}_{2}}$,$m_{\widetilde{b}_{R}}$,$m_{\widetilde{b}_{L}}$,$m_{\widetilde{u}_{R}}$,$m_{\widetilde{u}_{L}}$,$m_{\widetilde{d}_{R}}$,$m_{\widetilde{d}_{L}}$) occurs in the range of 854-1088 GeV. The minimum boundary on $m_{\widetilde{t}_{1}}$ is about 520 GeV.}
	\label{fig:Wedge_Masses}
\end{figure*}

A primary concern accompanying the release of a substantially upgraded quantity of data
which probes post-SM physics with unprecedented energy resolution is whether or not one's preferred model has
survived the search.  A secondary, though entangled, question is whether one's preferred model may in fact have
been glimpsed, even perhaps faintly, peeking over the background.  We first attempted to mimic an earlier generation
of the cuts favored by the CMS collaboration~\cite{Khachatryan:2011tk,PAS-SUS-09-001} in Refs.~\cite{Maxin:2011hy,Li:2011hr}.
The detailed selection criteria employed are summarized in Table~(II) of Ref.~\cite{Maxin:2011hy} and the surrounding discourse
of Section~(IV).  For simplicity, both there and here, we describe the standard intermediate jet count searches in shorthand
to be of the ``CMS'' style, although our discussion is obviously of equal relevance to the sister ATLAS detector.
We presently implement a few minor updates to the original ``CMS'' cuts, designed to mirror the most recently presented collaboration
physics analysis summary~\cite{PAS-SUS-11-003}.  In particular, we are adjusting the lower bounds of 350~GeV and 150~GeV
on the net scalar sum of transverse momentum $H_{\rm T}$ and the missing transverse energy $H_{\rm T}^{\rm miss}$ to 375~GeV and
100~GeV, respectively.  Although the CMS collaboration now more broadly advocates a larger quantity of individual $H_{\rm T}$ bins,
starting at $275$, with individually tuned thresholds on the transverse momentum $p_{\rm T}$ per jet in the lower two bins,
a useful presentation of the observed signal and background calibration is presented in Section~(2.1) of Ref.~\cite{PAS-SUS-11-003}
with the single $H_{\rm T} \ge 375$~GeV cut, which we shall for simplicity follow.  Additionally, the maximal pseudo-rapidity
$\eta$ of the leading jet has been raised from 2.0 to 2.5, although the threshold of 3.0 for sub-leading jets remains fixed.

Other selection criteria generally remain in place, including a lower bound of 50~GeV for the transverse momentum
$p_{\rm T}$ of each hard jet, an upper bound of 0.9 on the the electromagnetic fraction per jet
${(1+ {\rm had}/{\rm em})}^{-1}$ of calorimeter deposition, an upper bound of 1.25 on the missing energy ratio $R(H_{\rm T}^{\rm miss})$
of hard ($p_{\rm T} \ge 50$~GeV) to soft ($p_{\rm T} \ge 30$~GeV) jets, and upper bounds of $10$~GeV and $25$~GeV, respectively, for
the transverse momentum of isolated light leptons (electron, muon) and photons.  The latest CMS report more specifically describes
$R(H_{\rm T}^{\rm miss})$ as the ratio of missing energy computed from only hard jets to the same as computed from the calorimeter tower
estimate, but the intended function as a blockade against substantial cumulative failures of per jet limit
on transverse momentum is identical, and we expect comparable outcomes.  As a safeguard, we also secondarily filter with
regards to the missing energy reported natively by {\tt PGS4}.  The ``biased''  $\Delta \phi^*$ statistic, designed to help distinguish actual
missing energy signals from detector mismeasurements, remains disabled.  The CMS collaboration does employ
$\Delta \phi^*$ in conjunction with filters on proximity of the likely source of missing energy to masked regions of the electromagnetic
calorimeter. We do not attempt to replicate either this behavior or a similar treatment of the CMS barrel-endcap gap.
For our immediate comparison with Section (2.1) of the CMS source document~\cite{PAS-SUS-11-003}, the minimal jet count of three is
temporarily suspended, with two-jet events also allowed.  A lower bound of 0.55 is generally imposed on $\alpha_{\rm T}$, which was like
$\Delta \phi^*$, also devised to help isolate legitimate missing transverse energy, although this cut is also sometimes also relaxed,
either for the sake of comparison or for direct histogram binning against $\alpha_{\rm T}$ itself.  The role of the $\alpha_{\rm T}$ cut in
ultra-high jet multiplicity searches is of particular interest to our group, and we shall revisit the topic in detail in Section~(\ref{sct:less}).

In the figures described in this section, we reprint the signal and CMS Preliminary SM background statistics presented in Ref.~\cite{PAS-SUS-11-003},
overlaying our Monte Carlo simulation of each No-Scale \fsu5 benchmark~\cite{Li:2011in,Li:2011gh,Li:2011rp}, as well as the CMSSM benchmark
``Snowmass Points and Slopes'' SPS3~\cite{Allanach:2002nj}.  Consult Table~(\ref{tab:compareN}), which is printed following in Section~(\ref{sct:tenfold}),
for the map between the alphabetical model index and the corresponding LSP, gaugino, and vector-like masses $m_{\rm LSP}$,
$M_{1/2}$ and $M_{\rm V}$.
Our simulation was performed using the {\tt MadGraph}~\cite{Stelzer:1994ta,MGME} suite, including the standard
{\tt MadEvent}~\cite{Alwall:2007st}, {\tt PYTHIA}~\cite{Sjostrand:2006za} and {\tt PGS4}~\cite{PGS4} chain, with
post-processing performed by a custom script {\tt CutLHCO}~\cite{cutlhco} (available for download) which implements the desired
cuts, and counts and compiles the associated net statistics.  All 2-body SUSY processes have been included in our simulation,
which follows in all regards the procedure detailed in Ref.~\cite{Maxin:2011hy}.  Our SUSY particle mass calculations have been
performed using {\tt MicrOMEGAs 2.1}~\cite{Belanger:2008sj}, employing a proprietary modification of the
{\tt SuSpect 2.34}~\cite{Djouadi:2002ze} codebase to run the RGEs.
Most often, we oversample the Monte Carlo and scale down to the required luminosity, which can
have the effect of suppressing statistical fluctuations.

Figure~(\ref{fig:N_Jets}) depicts the event count per jet multiplicity, as based on Figure~(1a) from Section~(2.1) of
Ref.~\cite{PAS-SUS-11-003}.  It is most important to note that the $\alpha_{\rm T}$ cuts have been bypassed in this plot.  The logarithmic
event scaling generally allows one to disregard component signals which exhibit substantial visual separation below the leading term.
The SM expectation is, for the most part, a very good fit to the data in this plot, although there is an intriguing excess observed
above the SM background in the nine jet bin.  Visually, the central value of this
excess appears to be about four events, and allowing for extremities in the variation of both the expectation and the observation, it seems
that a range of one to seven events might be considered marginally consistent.  It seems in particular that the lighter portion of the No-Scale
\fsu5 model space, as represented by the lower alphabetically assigned indices (A-E), might be considered to overproduce above the observation.
Above this extreme, there is a healthy swath of parameter space that seems
quite nicely capable of accounting for the observation, carrying labels (F) and greater.  We have given graphical emphasis to two points, $\cal{F}$-J (bold green)
and $\cal{F}$-N (bold gold), with the green defining the estimated boundary of our surviving space subsequent to the present analysis, and the gold representing the most experimentally favored region consistent with the bare-minimal experimental constraints of~\cite{Li:2011xu} and both the $(b \rightarrow s\gamma)$ process and contributions to the muon anomalous magnetic moment $(g-2)_\mu$.
We note also that the CMSSM representative SPS3, shown in bold red dash, dramatically underperforms observations.  A similar result was clear in the
source document with regards to two tested CMSSM representatives, the benchmark models LM4 and LM6, with LM6 most similar to our SPS3 and faring the worse.
We can already see in this figure a considerably more rapid fatiguing of the SM and CMSSM signals with increasing jet count than is experienced
by any of the No-Scale \fsu5 candidates.  The eight and ten jet bins seem to consistently disfavor a similar swath of light models, although
there is again no constraint on the heavier models.  There were apparently no CMS observations in the eleven and twelve jet bins.  Since our favored
expectations are less than one event in each of these regions, we do not consider the statistics to be cautionary.

Figure set~(\ref{fig:4plex}) makes a similar treatment of Figures~(1a,1c,2a and 2b) of Ref.~\cite{PAS-SUS-11-003}.  The $\alpha_{\rm T}$ cut is
disabled in the upper two plots, but active in the lower.  Whereas the No-Scale \fsu5 histograms merge into something of a continuum, one is 
less visually interested in tracking the behavior of single model elements than delineation of the model space boundary.  We observe no
constraint from either of the upper figures.  The histogram in $\alpha_{\rm T}$, however, does effectively demonstrate the rather more
rapid falloff of the No-Scale \fsu5 models with respect to larger values of this statistic.  Again, the SPS3 benchmark seems to behave most
similarly to the externally studied LM6 example.

The source documents of the lower two figures opt for a linear scaling which imposes some visual compression on the model space.
In the $\Delta \phi^*$ histogram at lower left, we observe relatively low tension relative against any of the presented models.  The most interesting
feature of this plot is in the angular range of 0.8 to 1.4 radians, where the uncertainty of the observed signal remains in reasonable proximity to the background, though one could envision a possible excess above the Standard Model.  Curiously, certain of the lighter models seem well suited to explaining an excess, perhaps at the cost of reducing somewhat the observed parity with the SM signal from 0.0 to 0.8 radians.  Our lighter bold-printed model, $\cal{F}$-J, produces about two events in this region, which is acceptable.  There are a few bins, namely at 0.7, 1.9 and 2.7 radians, where the lightest \fsu5 models, and/or the SPS3 benchmark might be said to exceed the measurement, although each of these data points also defy the trend established by their neighbors. Likewise, the LM4 model considered in the source document appears in some cases to overproduce.  We also mention the fact that the CMSSM models SPS3 (and similarly again LM6) feature a flatter angular distribution in $\Delta \phi^*$ than their \fsu5 counterparts, which tend to peak earlier, at about one radian, and then fall of faster; we will return to this point in Section~(\ref{sct:less}).

The lower right-hand element of Figure Set~(\ref{fig:4plex}) is a histogram in the ``effective mass'' $M_{\rm eff}$ of the event, which
we take to be a scalar sum of all measured transverse energy and missing transverse energy, for all event beam fragments, including soft
hadronic jets, leptons and photons.  One initial feature to catch the eye is a possible excess at around 1~TeV.  It seems that a post-SM
production in the wide range of about 1 to 15 events is minimally consistent.  Only the very lightest \fsu5 models seem to suffer here.  Again,
models in the vicinity of $\cal{F}$-J perform favorably. Curiously, the externally studied benchmark LM4 dramatically overproduces, not only here, but also for larger values of $M_{\rm eff}$, approaching 2~TeV.  Likewise, several of the previously disfavored lighter \fsu5 benchmarks, and to some degree also SPS3, overproduce in the range of about 1.1~TeV to 1.8~TeV.  We see no model under consideration which can account for the observed excess in the 480-720~GeV region, which might be taken to engender some skepticism of the measurements or SM modeling in this region.

We turn attention now to Figure~(\ref{fig:N_Jets_alpha_T}), adapted from Figure~(2c) of Ref.~\cite{PAS-SUS-11-003}, which is a second plotting
of measured events by jet count, but now with the $\alpha_{\rm T}$ cut firmly in place.  In the source document, model LM4 again dramatically overproduces across the range, and is strongly disfavored.  We see no strong tension between any of the models presently considered in this work, for jet counts of 2, 3, or 4.  At five jets, it seems that the post-SM contribution should not exceed five or six events.  This cuts rather hard
against the No-Scale \fsu5 model space, but in a manner similar to what has been already observed.  Benchmark $\cal{F}$-J sits near the edge
of the permissible value, and the heavier bold-printed benchmark, $\cal{F}$-N performs quite well.  Although it is difficult to see in this
figure, the SM tracks the observation closely here for six or more jets, with low uncertainties indicated, and it seems that any substantial
extraneous production should be disfavored. In this regard, at a count of six jets, the $\cal{F}$-J benchmark does experience some strain,
although the heavier models around $\cal{F}$-N are quite comfortable.  Our simulation of the CMSSM point SPS3 avoids conflict throughout this
metric, again tracking nicely with the published LM6 model.  We remark, not inadvertently, that Figure~(\ref{fig:N_Jets_alpha_T}), as compared to
Figure~(\ref{fig:N_Jets}) seems to favor models in which high jet events are heavily suppressed by the $\alpha_{\rm T}$ statistic.

In Figure~(\ref{fig:Wedge}), we display the full bare minimally constrained~\cite{Li:2011xu} parameter space of No-Scale \fsu5, including application
of a crosshatch over the lighter spectra which we would estimate to be excluded by the analysis of this section.  Although a large
number of our originally selected benchmarks were placed within this lighter region, as a relative portion of the full hyper-volume, the model reduction is surprisingly minor.  The vertically established exclusion boundary, which is appropriate insomuch as the sparticle spectrum is virtually independent of $\tan \beta$ and the top quark $m_{\rm t}$ or the vector-like $M_V$ masses, emphasizes the reductionism of parameterization which is inherent in No-Scale \fsu5. The SPS3 benchmark, which has been our standard control sample in recent work, was chosen in some part for it's viability in the face of earlier generation search results.  This accounts, of course, for it's comparably heavy LSP mass.  The crystal clear exclusion of this model, and likewise of CMSSM representatives in its not-so-immediate vicinity, therefore speaks quite strongly to the current trouble which the mSUGRA/CMSSM framework is experiencing.

The analysis of Ref.~\cite{PAS-SUS-11-003} concluded that squark and gluino masses of 1.25 TeV may now be excluded for values of the unified scalar mass at the GUT scale of $m_0$ $<$ 530 GeV. However, as depicted by the light stop squark, gluino, and right-handed up squark mass contours in Figure~(\ref{fig:Wedge_Masses}), the No-Scale \fsu5 model space does not share a similar fate. Our estimated lower bound on the single \fsu5 model parameter $M_{1/2}$ of 485 GeV translates into a lower bound on the gluino mass of 658-674 GeV for the entire model space, notably below the gluino mass constraint of 1.25 TeV established for the CMSSM for $m_0$ $<$ 530 GeV. In a similar vein, we choose the right-handed up squark $\widetilde{u}_{R}$ to graphically represent all the heavy squarks, and we likewise discover that the lower bound on $m_{\widetilde{u}_{R}}$ to be 956-1000 GeV, also substantially below the CMSSM squark mass constraint of 1.25 GeV for $m_0$ $<$ 530 GeV. When considering all the heavy squarks ($m_{\widetilde{t}_{2}}$,$m_{\widetilde{b}_{R}}$,$m_{\widetilde{b}_{L}}$,$m_{\widetilde{u}_{R}}$,$m_{\widetilde{u}_{L}}$,$m_{\widetilde{d}_{R}}$,$m_{\widetilde{d}_{L}}$), the lower mass constraints in the \fsu5 model space lie within 854-1088 GeV. It is important to note again that the light stop $\widetilde{t}_1$ mass is less than the gluino mass in \fsu5, apparent through the unique mass hierarchy $m_{\tilde{t}} < m_{\tilde{g}} < m_{\tilde{q}}$, thus the light stop mass limit will also be much less than that of the CMSSM, with an \fsu5 light stop minimum mass of about 520 GeV.

\section{Ultra-High Jet Multiplicities}

The ultra-high jet signal of No-Scale \fsu5 was first discussed in Refs.~\cite{Maxin:2011hy,Li:2011hr},
and is based upon the stable sparticle hierarchy $m_{\tilde{t}} < m_{\tilde{g}} < m_{\tilde{q}}$
of a light stop and gluino, much lighter than all other squarks, which exists across the full model space.
This mass hierarchy is expected to strongly generate back-to-back pair production events, {\it e.g.} of two heavy squarks
$\widetilde{q}\,\widetilde{\overline{q}}$, which proceed by a decay such as $\widetilde{q} \rightarrow q \widetilde{g}$, followed by (where we identify the virtual tops with parentheses) $\widetilde{g} \rightarrow \widetilde{t}_{1} (\overline{t}) \rightarrow t \overline{t} \widetilde{\chi}_1^{0}
\rightarrow W^{+}W^{-} b \overline{b} \widetilde{\chi}_1^{0}$ or $\widetilde{g} \rightarrow \widetilde{t}_{1} (\overline{t})
\rightarrow b \overline{t} \widetilde{\chi}_1^{+} \rightarrow W^{-} b \overline{b} \widetilde{\tau}_{1}^{+} \nu_{\tau}
\rightarrow W^{-} b \overline{b} \tau^{+} \nu_{\tau} \widetilde{\chi}_1^{0}$ (plus conjugate processes), which produces
up to eight primary jets, considering only the initial hard scattering cascade.  This count may be substantially increased by
the further cascaded fragmentation and hadronization into final state showers of photons, leptons, and mixed jets.

We have proposed a specific alternative to the leading ``CMS'' style intermediate jet count $(3,4,5)$
SUSY searches which is retuned for the emphasis of high and ultra-high $(9,10,11,12,\ldots)$ jet content signals.
The details of the proposed cutting strategy, which we will refer to by the ``ULTRA'' shorthand, are given in
Section~(IV) of Ref.~\cite{Maxin:2011hy}.  Although various optimizations in the limiting jet count and transverse momentum
$p_{\rm T}$ per jet were considered for the ULTRA cuts in the cited work, we settled there on, and maintain here, the aggressive combination
of ($p_{\rm T} > 20$, ${\rm jets} \ge 9$) as the baseline indicator of this selection procedure.  The simultaneous reduction of the
$p_{\rm T}$ jet threshold from $50$ relative to the CMS style cuts is an essential parallel ingredient in the prescription for revealing
an enhanced quantity of softer jets.  We do, however, maintain demands that the two leading jets carry $100$~GeV of transverse momentum
each.  Pseudo-rapidity cuts of $\eta \le 3$ for all jets, and of $\eta \le 2$ for the leading jet, are likewise enforced.
Since the ultra-high jet regime is greatly suppressed in the SM backgrounds, we have been able to relax
certain of the harsh cuts which are very effective for separating out the MSSM in intermediate jet searches, but which
simultaneously exert a costly attrition against the No-Scale \fsu5 signal.  Specifically, we effectively disable the cuts on
the electromagnetic fraction per jet, $\alpha_{\rm T}$ and the missing energy ratio $R(H_{\rm T}^{\rm miss})$.
These distinctions are critical, as we shall elaborate in Section~(\ref{sct:less}), to the success of ultra-high jet
multiplicity search strategies.  We emphasize again though, validating the old adage that less is sometimes more,
that the ultra-high jet blockade itself forms a sufficiently strong discriminant against both the SM and typical mSUGRA/CMSSM
attempts at a post-SM solution.  A cautious reconsideration of lower order SM background contributions will be the topic
of Section~(\ref{sct:bgs}).

In more recent work~\cite{Li:2011gh,Li:2011rp}, we have undertaken a massive Monte Carlo simulation of the collider-detector
response of the the full bare-minimally constrained~\cite{Li:2011xu} parameter space of No-Scale \fsu5, modeling both the present LHC
center of mass energy $\sqrt{s} = 7$~TeV and upgrades which extend into the bright future of a $\sqrt{s} = 14$~TeV beam.
In addition to a prediction for the absolute detection prospects of a given model under certain conditions,
there are two sensible modes of comparison which one may consider: A) it is possible to contrast competing models,
taking for example respective elements from the viable CMSSM and \fsu5 parameter spaces, while keeping the detector
cut methodology constant; B) one may highlight enhancements in the visibility of a single model which are attributable
to changes in the search methodology, and in particular to the data selection cuts which are employed.  Both comparative
modes have the advantage over an absolute study that particular uncertainties may be expected to cancel between the
respective analyses.  We have considered both modes A and B in the past, and will revisit both again here, although
our primary interest will be with regards to mode B, for the reasons described following.

Immediately, our interest has been greatly piqued by the announcements made during the recent conference season
that various analyses of LHC data in the high jet multiplicity $(6,7,8,\ldots)$ channels are currently underway.
Specifically, the ATLAS collaboration stated that they have amassed $1.23~{\rm fb}^{-1}$, and presented analyses
of up to $1.0~{\rm fb}^{-1}$ for the mono-jet and di-jet channels with no evidence of new physics.  Results for searches targeted at exotic models, in five or more jets with a per jet transverse momentum threshold of $p_{\rm T} \ge 50$~GeV, remain pending.  In Ref.~\cite{Collaboration:2011tq}, the ATLAS collaboration looked at multi-jet searches up to 6 jets, applying a very hard Cut of $p_{\rm T} \ge 60$~GeV, although the data sampling was only $2.4~{\rm pb}^{-1}$.

The CMS collaboration meanwhile presented new data at the level of $1.1~{\rm fb}^{-1}$ for mono and di-jet searches, but reverted to
the prior level of $35~{\rm pb}^{-1}$ for the study of events with $\ge 6$ jets from pair-produced gluinos.  While the stated target
of CMS searches into this high jet territory~\cite{Chatrchyan:2011cj} is the search for exotic R-parity violating SUSY models and Technicolor,
any basic experimental physics results are of course model-independent, and we still take notice.  A most interesting physics
analysis summary~\cite{PAS-SUS-11-003} has also been released by the CMS collaboration, focusing on application of the $\alpha_{\rm T}$
variable, but in some places distributing events binned by integral jet count from 2 up to 12; this is the report which formed the
centerpiece of our discussion in Section~(\ref{sct:1p1ifb}).

A key aim of the next section will then be distinguishing among various
high and ultra-high jet cutting scenarios, attempting to quasi-continuously span the range of possibilities, demonstrating
firstly that not all such high-jet selection cuts are created equal, and secondly outlining precisely what material
differences in outcome might be expected within the No-Scale \fsu5 model space by the application of the various cuts considered.
We will continue to replicate all simulation and analysis in the context of our CMSSM control
sample, the ``Snowmass Points and Slopes'' benchmark SPS3~\cite{Allanach:2002nj}.
To a certain extent, it might appear that we already have at least a portion of what we are asking for, insomuch
as the data for distribution of event count per jet has been released, and one may simply truncate the counts below
any certain desired high jet threshold.  Indeed, we have in the prior section demonstrated that the surviving No-Scale
\fsu5 model space may very effectively, and even precisely, account for certain anomalies observed in the presented data.  However, we suggest in particular that the additional filtering criteria imposed under the global CMS style cutting procedure has potentially decimated the ultra-high
$(9,10,11,12,\ldots)$ jet multiplicity content, and that a much stronger validation might thus be immediately possible..  This is a critical point on which we shall immediately begin to elaborate.
 
\section{The Ten-Fold Way\label{sct:tenfold}}

In Ref.~\cite{Li:2011gh}, we introduced a discovery index $N$, given by the following expression, where $S$ and $B$
are respectively the observed signal and background at some reference luminosity, conveniently chosen as $1~{\rm fb}^{-1}$. 
\begin{equation}
N = \frac{12.5\, B}{S^2} \times \left[ 1 + \sqrt{ 1 + {\left( \frac{2S}{5B} \right)}^2} \,\right]
\end{equation}
The value of $N$ is the relative luminosity factor by which both $S$ and $B$ should be scaled in order to
achieve a baseline value of five for the statistic of merit $S/\sqrt{B+1}$ for overall model visibility.
Note that in the limit of large backgrounds, where the ``$+1$'' safety factor is unnecessary, the discovery
index reduces to $N = 25 B / S^2$.  In that work, our interest was the absolute likelihood of discovery
of various elements of the bare minimally constrained~\cite{Li:2011xu} model space of No-Scale \fsu5.
Correspondingly, we published there in Table~(1) a detailed tabulation of the $S/\sqrt{B+1}$ statistic,
at $\sqrt{s} = 7$~TeV and $1~{\rm fb}^{-1}$ of luminosity, for each selected benchmark representative of
the model space, as well as the CMSSM control sample SPS3~\cite{Allanach:2002nj}.  We found a wide range of
possible values for $N$, ranging from 0.05 to 38.9 for LSP masses ranging respectively from about 75 to 190~GeV.
This was supplemented by Figure~(1) of the same work, wherein we showed event count histograms in the missing transverse energy
\htmiss for all models, for beam energies of $\sqrt{s} = (7,8,10,12~{\rm and}~14)$~TeV, and by Figure~(2), wherein we
plotted the discovery index $N$ for each of the same.  The discovery index was found to increase exponentially with the LSP
mass, and also to depend sharply on the beam energy, decreasing by a factor between about 50 and 3500
under a doubling of the beam energy from 7 to 14~TeV, for the lightest and heaviest spectra respectively.
In conjunction with the sister publication~\cite{Li:2011rp}, wherein specific mass limits were projected for
key SUSY partners under various collider conditions, we concluded that the model space of No-Scale \fsu5
would begin to be effectively probed under the current operating environment of $1~{\rm fb}^{-1}$ at
$\sqrt{s} = 7$~TeV, with a majority of the remainder accessible to an identical sampling of $1~{\rm fb}^{-1}$
after the $\sqrt{s} = 14$~TeV upgrade, and the entirety within reach of $10~{\rm fb}^{-1}$ of high energy data.
Of course, as we indicated at the time, these results were entirely dependent upon adoption of the
optimized Ultra-high jet cuts.

The purpose of the current section is a comparative study between various cut strategies, which attempt to span
the continuum between the canonical ``CMS'' style $\ge 3$ jet cuts which we modeled in Ref.~\cite{Maxin:2011hy}, and the
optimized $\ge 9$ ultra-high jet multiplicity strategy, again spanning present and future collider energies,
and the complete bare minimally constrained \fsu5 model space, as well as our standard CMSSM control sample.
We have chosen four representative cutting scenarios, including the two previously studied, {\it i.e}, the baseline
CMS style $\ge 3$ jet scenario with updates as described in Section~(\ref{sct:1p1ifb}), which we will here dub ``CMS:3'',
and the baseline ultra-high $\ge 9$ jet scenario, which will be referred to as ``ULTRA:9''.  To this set, we add two additional
methodologies, which represent each a change only in the jet count threshold, being similar to their source procedures in
all other regards.  From the CMS:3 cuts, we clone the new ``CMS:6'' selector, raising the jet threshold from 3 to 6,
and likewise lowering the threshold from 9 to 6 jets in the original ultra-high procedure, we introduce the new ``ULTRA:6'' selector.

It should be noted that in our absolute studies of ultra-high jet cutting scenarios up to this point, we have argued that the
$t \overline{t} + {\rm jets}$ background should constitute the leading SM competition, being in fact alone sufficient to
model the full standard model (SM) process contribution.  In the present
work, where we must consider also selection cuts which do not make heavy use of the net jet count to significantly
reduce the SM backgrounds, we should certainly cast a wider net.
In particular, for $\ge 6$ jet searches, it would seem prudent at least to also consider the $W^\pm$ processes.
For three or more jets, it seems that the gates are cast essentially wide open.
There is an interesting reference available~\cite{Baer:2010tk} which provides rather comprehensive background
modeling statistics for the full set of SM components up to six jets, but these are before the application of any cuts,
which makes a direct translation into our present numerical calculations rather difficult.
However, since the interest is again now purely comparative, one may na\"{\i}vely suspect that the common divisor
of a roughly proportional background rescaling will cancel to a good approximation, and that reuse of the
simplified Monte Carlo samples is valid.

This is in fact not strictly correct, as various components of the background will react in an individually differentiated manner
to the jet threshold cut parameter.  We will nevertheless proceed as if this approximation is correct for the
interim, following up in Sections~(\ref{sct:excavate} and \ref{sct:bgs}) with an attempt to provide reasonable estimates, based upon the most
recently presented LHC background measurements~\cite{PAS-SUS-11-003}, of how the inclusion of a more realistic SM modeling
might affect the absolute detection prospects of No-Scale \fsu5, especially under lower jet multiplicity selection thresholds.
Immediately, one may anticipate that if the unmodeled backgrounds, including vector boson production,
possibly also associated with top quark or jet production, pure QCD $(2,3,4)$ multijet events, and all $b \overline{b}$ associated
processes, are indeed unable to substantially penetrate the ultra-high jet multiplicity barrier, the differential visibility of No-Scale \fsu5
in the ULTRA cuts relative to the CMS cuts will improve.  Indeed, we mention preemptively that these considerations
could potentially escalate the advertised by an additional factor in the range of about five to ten.

\begin{table*}[htbp]
        \centering
        \caption{Seventeen representative points, labeled $\cal{F}$-(A \ldots Q), are selected for display
	from the No-Scale \fsu5 model space, satisfying the bare-minimal phenomenological constraints outlined in Ref.~\cite{Li:2011xu}.
        In addition, a CMSSM control sample is supplied, corresponding to the ``Snowmass Points and Slopes'' benchmark SPS3~\cite{Allanach:2002nj}.
	For this latter point, the $M_{\rm V}$ column is appropriated instead for $M_0$.
        Units of GeV are taken for the dimensionful parameters $m_{\rm LSP}, M_{1/2}, M_{\rm V}, m_{t}~{\rm and}~M_0$.
	The rightmost three columns display the ratio of the discovery index $N$ for the given model, under two different
	cutting methodologies.  The constant reference denominator is the Ultra-high multiplicity selection cut on at least
	9 jets (ULTRA:9).  This is compared to the traditional CMS style cuts with at least 3 (CMS:3) or 6 (CMS:6) jets, as well as a
	replica of the Ultra-high style cut with the jet threshold reduced to 6 (ULTRA:6), all else being equal.  The results are
	averaged over center-of-mass collision energies of $\sqrt{s} = 7,8,10,12~{\rm and}~14$~TeV.
        }
                \begin{tabular}{|c|c|c|c|c|c|c|c|c|} \hline
$       ~~{\rm Model} ~~    $&$    ~~~ m_{\rm LSP}~~~$&$  ~~~M_{1/2} ~~~$&$   ~~~~  M_{\rm v}~~~~$&$    ~~\tan\beta~~$&$   ~~~~ m_{\rm t}~~~~$&$
        ~~\frac{N{\rm (CMS:3)}}{N{\rm (ULTRA:9)}} ~~ $&$   ~~  \frac{N{\rm (CMS:6)}}{N{\rm (ULTRA:9)}} ~~ $&$  ~~  \frac{N{\rm (ULTRA:6)}}{N{\rm (ULTRA:9)}}   ~~     $      \\[+2pt]      \hline \hline
$       \cal{F}$-${\rm A}  $&$     74.8    $&$     385     $&$     3575    $&$     19.8    $&$     172.5   $&$     3.3     $&$     9.4     $&$     0.6     $       \\      \hline
$       \cal{F}$-${\rm B}  $&$     75.0    $&$     395     $&$     2075    $&$     19.7    $&$     172.5   $&$     3.7     $&$     9.4     $&$     0.7     $       \\      \hline  
$       \cal{F}$-${\rm C}  $&$     74.7    $&$     400     $&$     1450    $&$     19.5    $&$     173.7   $&$     4.1     $&$     10.0    $&$     0.7     $       \\      \hline
$       \cal{F}$-${\rm D}  $&$     75.0    $&$     410     $&$     925     $&$     19.4    $&$     174.4   $&$     4.8     $&$     10.4    $&$     0.8     $       \\      \hline  
$       \cal{F}$-${\rm E}  $&$     83.2    $&$     425     $&$     3550    $&$     20.4    $&$     172.2   $&$     4.6     $&$     8.5     $&$     0.9     $       \\      \hline
$       \cal{F}$-${\rm F}  $&$     83.1    $&$     435     $&$     2000    $&$     20.1    $&$     173.1   $&$     5.8     $&$     9.1     $&$     1.0     $       \\      \hline  
$       \cal{F}$-${\rm G}  $&$     82.8    $&$     445     $&$     1125    $&$     19.9    $&$     174.4   $&$     6.8     $&$     10.0    $&$     1.1     $       \\      \hline
$       \cal{F}$-${\rm H}  $&$     92.2    $&$     465     $&$     3850    $&$     20.7    $&$     172.2   $&$     7.5     $&$     8.3     $&$     1.3     $       \\      \hline  
$       \cal{F}$-${\rm I}  $&$     92.2    $&$     475     $&$     2400    $&$     20.6    $&$     173.1   $&$     8.6     $&$     8.7     $&$     1.4     $       \\      \hline
$       \cal{F}$-${\rm J}  $&$     92.1    $&$     485     $&$     1475    $&$     20.4    $&$     174.3   $&$     9.8     $&$     9.2     $&$     1.6     $       \\      \hline  
$       \cal{F}$-${\rm K}  $&$     100.5   $&$     505     $&$     3700    $&$     21.0    $&$     172.6   $&$     10.4    $&$     7.6     $&$     1.9     $       \\      \hline
$       \cal{F}$-${\rm L}  $&$     100.2   $&$     510     $&$     2875    $&$     21.0    $&$     174.1   $&$     11.4    $&$     8.0     $&$     2.0     $       \\      \hline  
$       \cal{F}$-${\rm M}  $&$     100.0   $&$     520     $&$     1725    $&$     20.7    $&$     174.4   $&$     12.0    $&$     8.0     $&$     2.1     $       \\      \hline
$       \cal{F}$-${\rm N}  $&$     108.8   $&$     560     $&$     1875    $&$     21.0    $&$     174.4   $&$     15.7    $&$     7.1     $&$     2.7     $       \\      \hline  
$       \cal{F}$-${\rm O}  $&$     133.2   $&$     650     $&$     4700    $&$     22.0    $&$     173.4   $&$     20.7    $&$     5.8     $&$     4.1     $       \\      \hline
$       \cal{F}$-${\rm P}  $&$     155.9   $&$     750     $&$     5300    $&$     22.5    $&$     174.4   $&$     10.3    $&$     2.2     $&$     4.3     $       \\      \hline  
$       \cal{F}$-${\rm Q}  $&$     190.5   $&$     900     $&$     6000    $&$     23.0    $&$     174.4   $&$     7.5     $&$     1.7     $&$     4.3     $       \\      \hline
$    ~~   \cal{F}$-${\rm Average} ~~$&$100.8$&$    513     $&$     2859    $&$     20.7    $&$     173.5   $&$     8.7     $&$     7.9     $&$     1.8     $       \\      \hline \hline
$     ~~  {\rm SPS3}  ~~    $&$     161.7   $&$    400     $&$     M_0 = 90$&$     10.0    $&$     175.0   $&$     0.11    $&$     0.72    $&$     0.26    $       \\      \hline
                \end{tabular}
                \label{tab:compareN}
\end{table*}

We present in Table~(\ref{tab:compareN}), for each single model taken in turn, the ratio representing
the discovery index $N$ which applies under one of three competing selection cut scenarios, divided
by the value of $N$ which is achieved in our baseline ultra-high jet paradigm.
The incredibly strong dependence of the discovery index on the LSP mass and on the collider
energy which was described previously is found to be substantially tamed by the chosen ratio.
The residual distinctions between various model elements are found to be stronger, on balance,
than the distinctions arising from escalation of the beam energy.  We therefore average over
the latter in the primary table.  In Table~(\ref{tab:compareE}), we do explicitly demonstrate
the energy dependence by averaging instead over the model, although a wholly systematic effect is
not readily apparent.

\begin{table}[htbp]
        \centering
        \caption{Discovery index ratios are presented in a fashion similar to Table~(\ref{tab:compareN}), but averaged over the model space, and
	displayed independently per each collider energy. 
        }
                \begin{tabular}{|c|c|c|c|c|} \hline
$     ~~  {\rm Model}	~~$&$	~~\sqrt{s}  ~~   $&$
 ~\frac{N{\rm (CMS:3)}}{N{\rm (ULTRA:9)}}~ $&$  ~\frac{N{\rm (CMS:6)}}{N{\rm (ULTRA:9)}}~ $&$ ~\frac{N{\rm (ULTRA:6)}}{N{\rm (ULTRA:9)}}~     $       \\[+2pt]      \hline \hline
\multirow{5}{*}{$\cal{F}$-$SU(5)$}
&$     7       $&$     7.9     $&$     8.1     $&$     1.9$       \\      \cline{2-5}
&$     8       $&$     9.9     $&$     9.7     $&$     2.1$       \\      \cline{2-5}
&$     10      $&$     9.5     $&$     6.5     $&$     2.0$       \\      \cline{2-5}
&$     12      $&$     8.2     $&$     6.7     $&$     1.7$       \\      \cline{2-5}
&$     14      $&$     7.9     $&$     8.2     $&$     1.5$       \\      \hline \hline
\multirow{5}{*}{${\rm SPS3}$}
&$     7       $&$     0.04    $&$     0.57    $&$     0.16$       \\      \cline{2-5}
&$     8       $&$     0.08    $&$     1.02    $&$     0.21$       \\      \cline{2-5}
&$     10      $&$     0.13    $&$     0.50    $&$     0.32$       \\      \cline{2-5}
&$     12      $&$     0.14    $&$     0.62    $&$     0.30$       \\      \cline{2-5}
&$     14      $&$     0.19    $&$     0.87    $&$     0.30$       \\      \hline
                \end{tabular}
                \label{tab:compareE}
\end{table}

We can see clearly from Table~(\ref{tab:compareN}) that the traditionally styled CMS:3 cuts are actually
quite appropriate for elucidating the signal of the CMSSM, in fact some factor of order 10 more
appropriate than the most severe ULTRA variety of cuts.  Conversely, the ULTRA:9 cuts are definitively
the best tested cut for application to the No-Scale \fsu5 models, and likewise some factor of order 10
superior to the CMS:3 variety, for this purpose.  However, this analysis, by design is purely comparative,
and reveals nothing about the absolute visibility of either model.  For this purpose, we must either directly
access the discovery index $N$, accepting also the associated complication of the SM background, or
compose again an appropriate ratio.  We will opt to evaluate the ratio of the discovery index $N$
for the No-Scale \fsu5 models divided by the CMSSM discovery index.  The backgrounds literally cancel in
this case, at least in the large background limit, and the result reduces essentially to a ratio of
inverse signals-squared.  It is difficult, though, to establish a rationale for which points of the respective
model spaces are to be compared, as the expected signal varies dramatically with the LSP mass. Purely
as a matter of individual discretion, we will partition the No-Scale \fsu5 benchmarks into light
($m_{\rm LSP} < 95~{\rm GeV}$), medium ($95~{\rm GeV} \le m_{\rm LSP} < 135~{\rm GeV}$), and heavy
($135~{\rm GeV} \le m_{\rm LSP}$) groups, and give ratios in Table~(\ref{tab:compareA}) for the
CMSSM benchmark SPS3 discovery index in comparison to the average \fsu5 discovery index, for each mass
grouping, and for each beam energy.

\begin{table}[htbp]
        \centering
        \caption{The ratio of absolute visibility of No-Scale \fsu5 versus the CMSSM benchmark SPS3 is
	tabulated for various collider energies and various cut methodologies.  The elements of the \fsu5
	model space are partitioned for this comparison into light, medium and heavy groupings, according
	to the mass of the LSP.
        }
                \begin{tabular}{|c|c|c|c|c|} \hline
$     ~\sqrt{s}~$&$    ~{\rm CMS:3}~$&$  ~{\rm CMS:6}~$&$  ~{\rm ULTRA:6}~$&$  ~{\rm ULTRA:9}~   $       \\      \hline\hline
\multicolumn{5}{|c|}{$m_{\rm LSP} < 95~{\rm GeV}$ \quad ($\cal{F}$-A \ldots $\cal{F}$-J)} \\ \hline
$       7       $&$     1.6     $&$     19.5    $&$     37.5    $&$     317.8   $       \\      \hline
$       8       $&$     1.7     $&$     16.3    $&$     30.3    $&$     176.5   $       \\      \hline
$       10      $&$     1.8     $&$     5.5     $&$     26.2    $&$     88.8    $       \\      \hline
$       12      $&$     1.7     $&$     5.8     $&$     22.7    $&$     74.9    $       \\      \hline
$       14      $&$     1.8     $&$     5.4     $&$     19.1    $&$     59.6    $       \\      \hline
$~{\rm Average}~$&$     1.7     $&$     10.5    $&$     27.1    $&$     143.5   $       \\      \hline\hline
\multicolumn{5}{|c|}{$95~{\rm GeV} \le m_{\rm LSP} < 135~{\rm GeV}$ \quad ($\cal{F}$-K \ldots $\cal{F}$-O)} \\ \hline
$       7       $&$     0.0     $&$     1.3     $&$     0.6     $&$     14.4    $       \\      \hline
$       8       $&$     0.1     $&$     1.8     $&$     0.7     $&$     15.1    $       \\      \hline
$       10      $&$     0.1     $&$     1.5     $&$     1.1     $&$     13.2    $       \\      \hline
$       12      $&$     0.1     $&$     1.7     $&$     1.3     $&$     13.4    $       \\      \hline
$       14      $&$     0.1     $&$     1.4     $&$     1.4     $&$     12.0    $       \\      \hline
$~{\rm Average}~$&$     0.1     $&$     1.5     $&$     1.0     $&$     13.6    $       \\      \hline\hline
\multicolumn{5}{|c|}{$135~{\rm GeV} \le m_{\rm LSP}$ \quad ($\cal{F}$-P \& $\cal{F}$-Q)} \\ \hline
$       7       $&$     0.00    $&$     0.00    $&$     0.00    $&$     0.01    $       \\      \hline
$       8       $&$     0.00    $&$     0.01    $&$     0.00    $&$     0.02    $       \\      \hline
$       10      $&$     0.00    $&$     0.04    $&$     0.00    $&$     0.06    $       \\      \hline
$       12      $&$     0.00    $&$     0.06    $&$     0.01    $&$     0.10    $       \\      \hline
$       14      $&$     0.00    $&$     0.06    $&$     0.01    $&$     0.14    $       \\      \hline
$~{\rm Average}~$&$     0.00    $&$     0.03    $&$     0.00    $&$     0.07    $       \\      \hline
                \end{tabular}
                \label{tab:compareA}
\end{table}

In absolute terms, Table~(\ref{tab:compareA}) demonstrates that the lighter elements of No-Scale \fsu5 are generally
about equally as visible as the SPS3 benchmark, under the CMS:3 cutting scenario.  This highlights the fact established in
Section~(\ref{sct:1p1ifb}) that these lighter models, along with huge swaths of the mSUGRA/CMSSM structure, have effectively
been disallowed by the first inverse femtobarn of LHC data.  The ULTRA:9 cuts are again about
ten times better for \fsu5, and about ten times worse for the CMSSM representative, but we can now see
that both differentials work in the same direction.  In other words, while the CMS cuts are indeed very
bad for No-Scale \fsu5, the lighter elements ($m_{\rm LSP} \le 95$~GeV) of this model can
still compete for visibility in standard searches when put up against sufficiently comparably heavier
elements of the CMSSM, insomuch as they enjoy the enhancements due to a light LSP.  However, we
of course only compare to a single CMSSM element, and one notably chosen in part for its persistent
(historical, though no longer) viability against experimental limits.  Conversely, applying the ULTRA:9 cuts to both models,
the balance tips decisively in favor of \fsu5, whose order of ten-fold absolute enhancement, combined with
the order of ten-fold SPS3 absolute suppression, nets a relative visibility advantage of up to 100 times.

It should be clearly remarked, though, that the key point here is not the advantage in discoverability over the SPS3 benchmark
itself, since as we have noted it makes very little sense to apply the ULTRA variety of cuts if one is attempting to probe the
CMSSM class of models.  A much more relevant comparison in this case is the advantage garnered by the ULTRA selection
cuts over the CMS selection cuts, holding the No-Scale \fsu5 model space constant.  As effectively demonstrated by
Tables~(\ref{tab:compareN},\ref{tab:compareE}), this advantage appears to stand remarkably stable, considering the
wide range of cross sections considered, near the advertised ratio of order ten.  Rather, the key importance of the
larger factor demonstrated in Table~(\ref{tab:compareA}) for comparison of \fsu5 against the CMSSM under constant
application of the ULTRA:9 cuts is its remarkable ability to differentiate competing explanations for the source of a
potential observed signal.  Strong differential performance under the ULTRA cutting methodology could then be construed
as strong evidence that an observed signal could be attributable to a model in the vicinity of the \fsu5 universe. 

Moving down into the medium weight models of Table~(\ref{tab:compareA}), we find that the balance shifts.
As should be expected, the visibility ratio of \fsu5 to SPS3 is up to 100 times higher under the ULTRA:9
cuts than under the CMS:3 cuts.  However, rather than parity in the model discovery ratio under the CMS:3 cuts,
we find an order of 10 suppression in the absolute visibility of \fsu5, and correspondingly, only an order of 10
enhancement under the ULTRA:9 cuts for the \fsu5 models in comparison to SPS3.  Of course, this is consistent
with the observation from Section~{\ref{sct:1p1ifb}} that the middle weight and heavy \fsu5 models have generally
not been ruled out by the CMS:3 cut strategy, while a majority of the lightweight classification has been, along with the SPS3 benchmark itself.
The heavy \fsu5 points are quite difficult to resolve, even under the ULTRA:9 cuts, as we have advertised
previously~\cite{Li:2011gh,Li:2011rp}, and will require fulfillment of the promised LHC energy upgrade.  The discovery
ratios in Table~(\ref{tab:compareA}) are all substantially less than one for this heavy category, even though, by LSP
mass, this is where the SPS3 benchmark belongs.  The situation is improved if the heaviest \fsu5 point is excluded,
but the difficulty in making a direct comparison between competing models is still underscored.

To this point, most of our discussion has focused on the extremes of the selection criteria range, namely the CMS:3 and ULTRA:9
methodologies, we conclude this section with some commentary on the two six jet selection cuts CMS:6 and ULTRA:6, which are in
some ways quite similar, {\it cf.}~Table~(\ref{tab:compareA}), and in some ways distinct, {\it cf.}~Table~(\ref{tab:compareN}).
The absolute discovery advantage of Table~(\ref{tab:compareA}) would tip in the favor of the light \fsu5 models for these
cases by a factor of order ten, and the favored middle weight models would exist at basic parity with the CMSSM sample.
This is consistent with and complementary to the observation from Table~(\ref{tab:compareN}) that the CMS:6 cuts are of roughly
equal standing with the ULTRA:9 cuts from the CMSSM perspective, although of order ten times worse from the \fsu5 perspective,
performing in this case comparably to the CMS:3 cuts.  Likewise, the ratio ULTRA:6/ULTRA:9 is numerically larger than one, but still
of order one for the \fsu5 space, and comparatively on the order of ten times smaller when applied to the SPS3 benchmark.
For the heavier models, the visibility of the \fsu5 models relative to the SPS3 benchmark, although clearly much weaker
as a fixed metric, retains the same relative scaling which is apparent for the lighter models, being an order of ten times
stronger for either the six jet cut than the CMS:3 cuts, and an order of ten times weaker for either the six jet cut than the ULTRA:9 cut.

If the LHC detector collaborations should shortly release an analysis similar to the selection cuts described here as
CMS:6, this would certainly be a step in the right direction, from our point of view.  However, it is not immediately
clear that this step would in isolation offer a significant improvement over the picture established by the integral
jet binning which has already been provided, for example by the CMS collaboration plots reprinted in
Figures~(\ref{fig:N_Jets},\ref{fig:N_Jets_alpha_T}).  Taken in conjunction, what Tables~(\ref{tab:compareN} and \ref{tab:compareA})
seem to imply is that, while critical to the detectability of No-Scale \fsu5, jet count itself is not the only criterion of
interest for detection of the No-Scale \fsu5 models.  In particular, it must be emphasized that not all six-jet (or nine
jet or etc.) searches are comparable simply at face value.  The question of how and why the two six-jet selection cuts are
indeed in some ways so different is the focus of Section~(\ref{sct:less}).

\section{Less Becomes More\label{sct:less}}

In this section we emphasize distinctions between the CMS and ULTRA selection cut
philosophies which go beyond the obvious shift in the minimal jet count threshold, casting additional
light on the dramatic differences which were observed between the two trial six-jet selection cuts discussed
in Section~(\ref{sct:tenfold}).  Specifically, the remaining differences involve the minimal transverse
momentum required for classification as a jet, and a disabling of cuts on the electromagnetic
fraction ratio, the ratio $R(H_{\rm T}^{\rm miss})$ of hard to soft jets, and the $\alpha_{\rm T}$ statistic.

Averaged across the model space, we find that the CMS:3 cuts at $\sqrt{s} = 7$~TeV have a filtering efficacy of 97.6\% against the \fsu5 signal.
This is somewhat disappointing, considering that the sample is composed entirely of the targeted SUSY events.  It is an awful lot of gold
washing over the pan and down the stream.  By comparison, only 91.0\% of the SPS3 benchmark was so cut.  This suggests strongly that a much more
efficiently graded tier of filters must be employed to better differentiate the sand and waste from the few precious nuggets which may pass our way.
Digging deeper into the cutting statistics, we find that a full 71.2\% of events are cut for not producing at least three jets which satisfy all
selection criteria, including the per jet transverse momentum threshold of $50$~GeV.  Only 1.1\% of events are cut for possessing jets which pass the hard momentum threshold while failing either the electromagnetic fraction or pseudo-rapidity limits. The cuts on the leading jet pseudo-rapidity and the transverse momentum of the two leading jets affect 53.8\% and 72.9\% of all events respectively.  The cuts on jet missing energy and the scalar sum of jet transverse momentum net capture rates of 71.1\% and 72.1\%.  Limits on isolated energetic leptons and photons affect 22.5\% and 1.5\% of
events, while the ratio $R(H_{\rm T}^{\rm miss})$ acts on 3.9\%.  The dominant cut, however, is that on $\alpha_{\rm T}$, which is active on
a total of 92.9\% of all events.  Even more suggestively, $\alpha_{\rm T} \ge 0.55$ represents the only cut which has a substantial efficacy
as the sole active exclusion on a large number of events, accounting in isolation for the rejection of a full 10.3\% of the sample.

It may seem counter intuitive to abandon tried and true techniques which have proven so beneficial for reduction of the SM background
against intermediate jet multiplicity events, but we have observed in simulation that, worse even than simply failing to effectively
differentiate the ultra-high jet signal, certain of these well known markers may actually preferentially indicate against it~\cite{Maxin:2011hy}.
It is worth recalling here that the $\alpha_{\rm T}$ statistic was originally devised for di-jet processes,
and later adapted to multi-jet events by the artful assemblage of two optimized pseudo-jets from the
full set of tracks.  Its intrinsic relevance for the scaling up to ultra-high jet processes may then be held in some doubt.
Indeed, we have observed, {\it cf.} the upper right plot of Figure Set~(\ref{fig:4plex}), a rather rapid decline in \fsu5
events beyond the value of $\alpha_{\rm T} = 0.5$.  We remind the reader also that the feature of rapid attrition of high jet counts under the
$\alpha_{\rm T} \ge 0.55$ selection cut was particularly relevant to the success of \fsu5 in the analysis of Figure~(\ref{fig:N_Jets_alpha_T}).
In order to better account for these interrelated observations, we now present the formal definition of the jet-based missing energy
$H_{\rm T}^{\rm miss}$, and also of $\alpha_{\rm T}$ for multijet events.  Likewise, we recall that
$H_{\rm T} \equiv \sum_{\rm jets} \left| {\vec{p}}_{\rm T} \right|$ is the simple scalar sum across the magnitudes of the transverse
momenta of all hard jets.
\begin{equation}
H_{\rm T}^{\rm miss} \equiv
\sqrt{{\left( \sum_{\rm jets} p_{\rm T} \cos \phi \right)}^2 + {\left( \sum_{\rm jets} p_{\rm T} \sin \phi \right)}^2}
\label{eq:HTmiss}
\end{equation}

\begin{equation}
\alpha_{\rm T} \equiv \frac{1}{2} \left\{ \frac{ 1- \left( \Delta H_{\rm T}^{\rm MIN} / H_{\rm T} \right)}{1 - {\left( H_{\rm T}^{\rm miss} / H_{\rm T} \right)}^2} \right\}
\label{eq:alpha_T}
\end{equation}
The graphical interpretation of Eq.~(\ref{eq:HTmiss}) is that of a vector sum of directed transverse momenta, resulting in an open
polygonal shape, whose missing leg gives the magnitude of the missing energy signal.
In Eq.~(\ref{eq:alpha_T}), $\Delta H_{\rm T}$ is the (positive) difference in the net scalar transverse momentum between two arbitrarily
partitioned groupings of the surviving jets.  All such possible combinations of pseudo jets are considered, and the minimal value of
$\Delta H_{\rm T}$ is employed.  If there is no mismeasurement or true missing energy, the value
of $\alpha_{\rm T}$ will just be $1/2$.  For energy magnitude mismeasurements of otherwise anti-parallel (pseudo) jet pairs,
subtraction of the nonvanishing scalar difference $\Delta H_{\rm T}$ will tend to drive $\alpha_{\rm T}$ below the midline.  Note that
in this case $\Delta H_{\rm T} = H_{\rm T}^{\rm miss}$, but the squaring of the small (typically $\ll 1$) factor in the denominator renders it
less significant.  Conversely, genuine missing energy, as manifest in the departure from (pseudo) jet anti-parallelism, will imbalance the vector sum
within the factor $H_{\rm T}^{\rm miss}$ of the denominator more so than the simple magnitude difference $\Delta H_{\rm T}$, tending to create a
contrasting elevation in $\alpha_{\rm T}$ above one-half.

The central point which explains the failure of the $\alpha_{\rm T}$ statistic at ultra-high jet multiplicities seems to be that a large
availability of relatively soft jets allows for such a wide array of pseudo-jet combinations that it becomes quite likely that a
balanced scalar sum $\Delta H_{\rm T}^{\rm MIN} \simeq 0$ might be achieved, creating an overall suppression of the $\alpha_{\rm T}$ value.
A very similar attribution may be made for the reason of the (possibly quite experimentally favorable) anomalous bias of $\Delta \phi^*$
toward zero in the \fsu5 models, as depicted in the lower left-hand plot of Figure Set~(\ref{fig:4plex}).  Although $\Delta \phi^*$ is not expressly
activated in any of the selection criteria described in this work, it remains a statistic of common use and relevance which is designed to
establish whether poor scaling of a single jet measurement might be responsible for a false missing energy signature.  Specifically, for each
surviving jet in turn, $\Delta \phi^i$ registers the absolute azimuthal angle in the range $(0,\pi)$ which separates the transverse momentum
vector of the $i^{th}$ jet from the negation of the directional imbalance which arises by omitting that jet from the vector transverse momentum sum.
The minimal such value, denoted with the index ``$*$'' is the one reported.  If a single jet mismeasurement is indeed dominantly responsible for
a false missing energy signal, then $\Delta \phi^*$ should register close to zero.  However, again, given a wide selection of randomly oriented
jets, it becomes in fact quite likely that the angular orientation of at least one such member might be sufficiently well azimuthally aligned
with the true missing energy track that its rescaling could apparently rebalance the event.  Visually, we are no longer stretching one leg of
an open triangle to make it close to zero length under vector addition and so eliminate the missing energy signal. We are instead perhaps
stretching one leg of an open decagon!  The $\alpha_{\rm T}$ selector (and likewise $\Delta \phi^*$) appears simply unsuited for ultra jet
searches, and is responsible for a catastrophic signal suppression for the reason described.

If our goal is excavation of the supersymmetric No-Scale \fsu5 signal out of the Standard Model rubble, then a
critical simultaneous complement to the increased jet multiplicity threshold and elimination of the $\alpha_{\rm T}$ cut
seems to be a lowering of the required transverse momentum per jet.  Indeed, even without the $\alpha_{\rm T}$ cut applied,
Figure~(\ref{fig:N_Jets}) demonstrates a paucity of $\ge 9$ jet events, a fact which we attribute primarily to the large
per-jet transverse momentum threshold of $p_{\rm T} \ge 50$.  We have again adopted the aggressive cuts $p_{\rm T} > 20$ and  ${\rm jets} \ge 9$ as the baseline ULTRA
selection procedure.  There is very recent documentation from the ATLAS collaboration~\cite{Aad:2011gn} which finds that soft jets below about 20~GeV are not modeled well, with simulations diverging from the actual data. However, above this threshold, the correspondence with data is reported to be quite satisfactory; we find this to be a most affirmative result for our preferred level of selection, which was in some sense designed to target a lower cusp of reasonable efficacy. It is certainly true that a softening of the per-jet momentum threshold will admit into the calculus a significant decentralization from the initial hard scattering intermediate Feynman diagrams. Nevertheless, the basic intuition that fewer hard jets in the early parton level diagrams will yield a correspondingly smaller count of final state soft jets is well confirmed by the Monte Carlo, and we have demonstrated not only a readily detectable signal for No-Scale $\cal{F}$-$SU(5)$ above the SM background, but also a clear differentiation
between No-Scale $\cal{F}$-$SU(5)$ and a typical competing post-SM scenario.  The unique SUSY mass hierarchy of No-Scale $\cal{F}$-$SU(5)$,
which we have not found replicated by any models of the CMSSM variety, convinces us of the broad generality of these conclusions.

Aware of these considerations, it is worth remarking that some caution is in order to ensure that this same principle in reverse
does not undermine the missing energy measurement itself at high jet multiplicities.  Again, our Monte Carlo indicates,
despite a strong likelihood that the prior concern does play some role in the simulation, that the ULTRA:9 cuts based on jet multiplicity,
pseudo rapidity per jet, transverse momentum $p_{\rm T}$ per jet, the net jet energy $H_{\rm T}$, and the net jet missing energy $H_{\rm T}^{\rm miss}$
do provide striking differentiation of the No-Scale \fsu5 models above the competing SM background, and also with respect to competing
mSUGRA/CMSSM proposals.  The need for extra suppression of backgrounds, a task frequently assigned to selections based on $\alpha_{\rm T}$,
is here circumvented by the simple observation that the ultra-high jet threshold itself accomplishes an extraordinarily robust suppression
of the Standard Model.  Indeed, it seems in this case, that less can amount to more.

\section{Excavating SUSY With A $1.1~{\rm fb}^{-1}$ LHC\label{sct:excavate}}

\begin{figure*}[htp]
        \centering
        \includegraphics[width=0.45\textwidth]{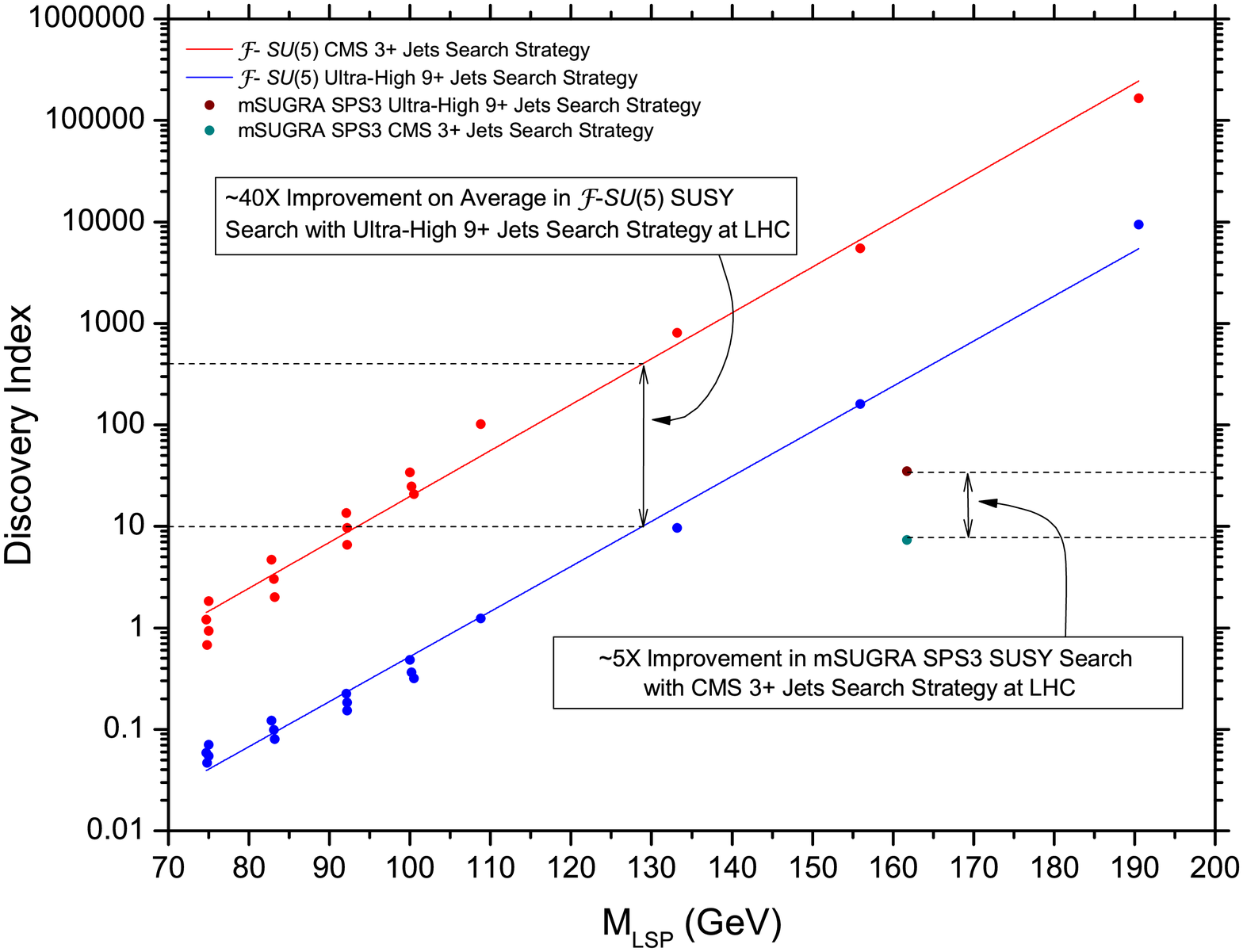}
	\hspace{0.05\textwidth}
        \includegraphics[width=0.45\textwidth]{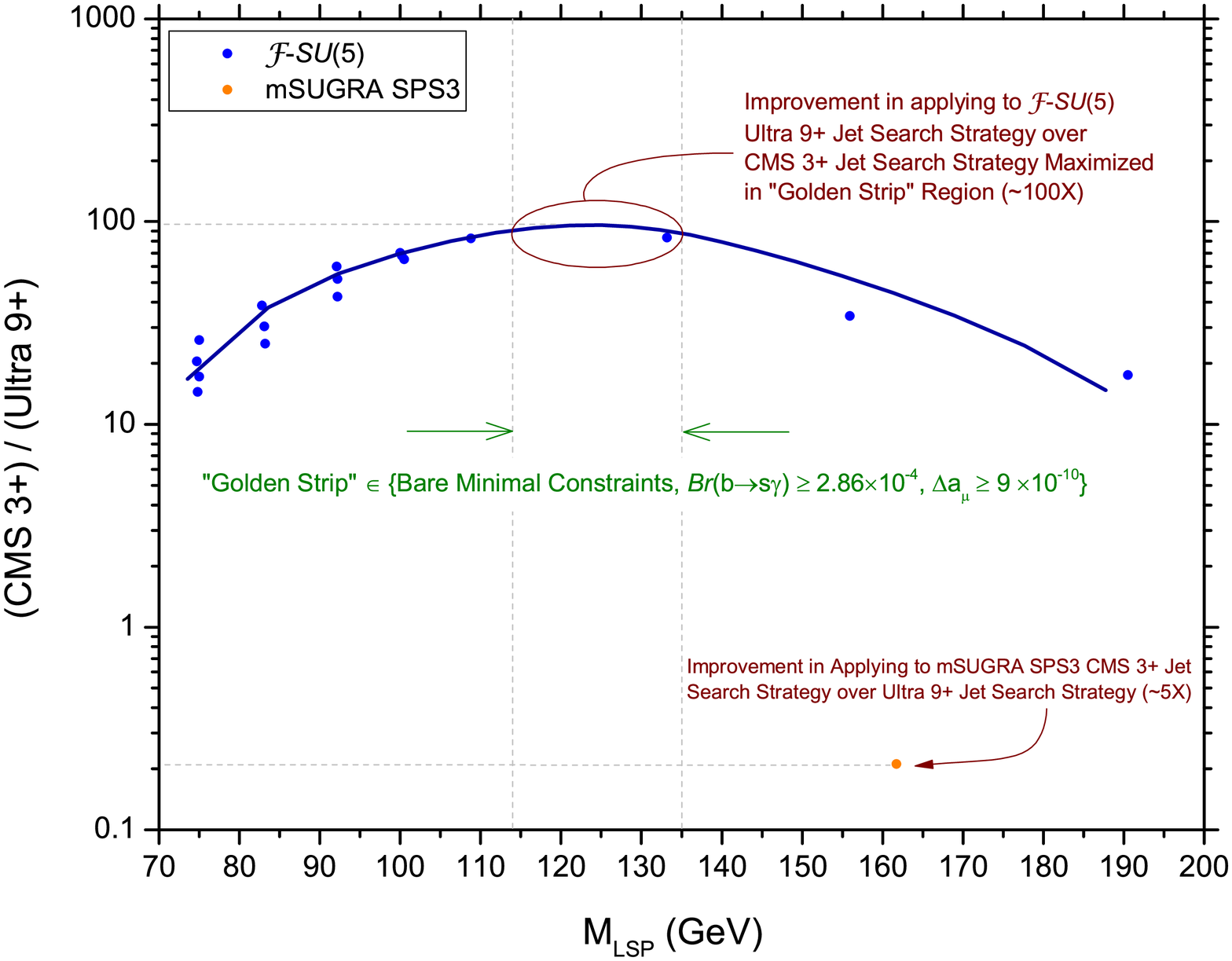}
        \caption{ The lefthand figure plots the absolute discovery index $N$ of the No-Scale \fsu5 model space, {\it i.e.}~the
	projected number of inverse femtobarns of luminosity which would need to be integrated in order to achieve a value
	of five for the signal visibility metric $S/\sqrt{B+1}$.  The backgrounds under the CMS:3 cuts
	are extracted from a CMS analysis~\cite{PAS-SUS-11-003} to a value of about 195 observations for $1~{\rm fb}^{-1}$ of data.  The
	backgrounds under the ULTRA:9 cuts are from our simulation of $t \overline{t} + {\rm jets}$ processes.  The right-hand
	figure demonstrates the distribution of the ULTRA:9 discovery advantage as a function of the LSP mass.  Curiously, the
	ratio is maximized nearest to the most experimentally favored region, which consists of satisfaction of the experimental limits
	on contributions to the anomalous magnetic moment of the muon $(g-2)_\mu$ and the branching ratio of the
	flavor changing neutral current process $(b \rightarrow s\gamma)$, over and above the bare-minimal constraints of~\cite{Li:2011xu}.  We define this region to constitute a new
	``Golden Strip'' within the bare-minimally constrained~\cite{Li:2011xu} No-Scale \fsu5 model space.
        } \label{fig:Ultra_Ratio}
\end{figure*}

The $t \overline{t} + {\rm jets}$ background is certainly not sufficient to model the detector response under the
CMS style three jet selection criteria, nor for the two considered six jet scenarios.  As elaborated in Section~(\ref{sct:1p1ifb}),
this had been a consideration in our decision
to focus in the present work on relative rather than absolute performance metrics, such as the comparative visibility advantage
garnered by the No-Scale \fsu5 model class under application of the ULTRA:9 cutting scenario versus the CMS:3 scenario.  We
argued that some cancellation in this ratio between additional contributions to the SM backgrounds should be expected, but
that the weaker performance at high jet multiplicities of those SM background components which have been neglected in simulation
implies that the reported ten-fold visibility advantage is in actuality a lower bound. In the present section, 
we attempt to quantify the factor by which the discovery advantage of the ULTRA selection cut criteria within the No-Scale
\fsu5 model space may be enhanced.

Table~(\ref{tab:compareNB}) represents a variation on the Table~(\ref{tab:compareN}) theme, again comparing single model
visibility relative to the application of various cuts.  We reduce the cutting scenarios, though, to only the CMS:3 and ULTRA:9 styles,
and the collider energy to only $\sqrt{s} = 7$~TeV.  The reason for this is that in Table~(\ref{tab:compareNB}) we wish to apply certain
more concrete values of the low jet background, taken directly from experiment.  For this purpose, we will extrapolate from graphical
backgrounds presented for $1.1~{\rm pb}^{-1}$ of CMS data~\cite{PAS-SUS-11-003}.
This translates in the present context to a value of about 195 observations for $1~{\rm fb}^{-1}$ of data.  We are assuming still adequate
background suppression under the ULTRA:9 cuts for all but our $t \overline{t} + {\rm jets}$ simulation.
Given actual data for the intermediate jet multiplicity cuts, we now add to the Table~(\ref{tab:compareN})
presentation a printing of the absolute discovery index $N$, {\it i.e.}~the projected number of inverse femtobarns of luminosity which
would need to be integrated in order to achieve a value of five for the signal visibility metric $S/\sqrt{B+1}$.

\begin{table}[htbp]
        \centering
        \caption{ An extension of the Table~(\ref{tab:compareN}) analysis is carried out for the seventeen \fsu5 benchmarks, as well as the
	CMSSM representative SPS3.  The collider energy represented is constant at $\sqrt{s} = 7$~TeV.  The backgrounds under the CMS:3 cuts
	are extracted from a CMS analysis~\cite{PAS-SUS-11-003} to a value of about 195 observations for $1~{\rm fb}^{-1}$ of data.  The
	backgrounds under the ULTRA:9 cuts are from our simulation of $t \overline{t} + {\rm jets}$ processes.  Since the backgrounds are
	better known here than was generally true in Table~(\ref{tab:compareN}), we are comfortable printing the absolute discovery index $N$,
	{\it i.e.}~the projected number of inverse femtobarns of luminosity which would need to be integrated in order to achieve a value
        of five for the signal visibility metric $S/\sqrt{B+1}$, for each model under the CMS:3 and ULTRA:9 cuts, in addition
	to the discovery ratio of the two procedures.  However, because of the strong model dependence, we suppress presentation of
	an average for these columns.  This data is represented in graphical form by Figure Set~(\ref{fig:Ultra_Ratio}).
        }
                \begin{tabular}{|c|c|c|c|} \hline
$       ~~{\rm Model} ~~    $&   ~$N$(CMS:3)~ & ~$N$(ULTRA:9)~ &$ ~\frac{N{\rm (CMS:3)}}{N{\rm (ULTRA:9)}}~   $    \\[+2pt]      \hline \hline
$       \cal{F}$-${\rm A}  $&$        0.7   $&$    0.05   $&$   14.5     $   \\      \hline
$       \cal{F}$-${\rm B}  $&$        0.9   $&$    0.05   $&$   17.2     $   \\      \hline
$       \cal{F}$-${\rm C}  $&$        1.2   $&$    0.06   $&$   20.5     $   \\      \hline
$       \cal{F}$-${\rm D}  $&$        1.8   $&$    0.07   $&$   26.1     $   \\      \hline
$       \cal{F}$-${\rm E}  $&$        2.0   $&$    0.08   $&$   25.0     $   \\      \hline
$       \cal{F}$-${\rm F}  $&$        3.0   $&$    0.10   $&$   30.5     $   \\      \hline
$       \cal{F}$-${\rm G}  $&$        4.7   $&$    0.12   $&$   38.7     $   \\      \hline
$       \cal{F}$-${\rm H}  $&$        6.6   $&$   0.15   $&$   42.7     $   \\      \hline
$       \cal{F}$-${\rm I}  $&$        9.6   $&$   0.18   $&$   52.2     $   \\      \hline
$       \cal{F}$-${\rm J}  $&$        13.6   $&$   0.23   $&$   60.1     $   \\      \hline
$       \cal{F}$-${\rm K}  $&$        20.8   $&$   0.32   $&$   65.2     $   \\      \hline
$       \cal{F}$-${\rm L}  $&$        24.7   $&$   0.36   $&$   67.9     $   \\      \hline
$       \cal{F}$-${\rm M}  $&$        34.2   $&$   0.49   $&$   70.3     $   \\      \hline
$       \cal{F}$-${\rm N}  $&$       102.3   $&$   1.23   $&$   82.9     $   \\      \hline
$       \cal{F}$-${\rm O}  $&$       808.5   $&$    9.7   $&$   83.5     $   \\      \hline
$       \cal{F}$-${\rm P}  $&$        5.49 \times10^3  $&$    160.1   $&$   34.3     $   \\      \hline
$       \cal{F}$-${\rm Q}  $&$        1.66 \times10^5  $&$    9.45\times10^3   $&$   17.5     $   \\      \hline
$    ~~   \cal{F}$-${\rm Average} ~~$&      --   &    --   &$   44.1     $   \\      \hline\hline
$     ~~  {\rm SPS3}  ~~    $&$        7.3   $&$    34.8   $&$   0.2     $   \\      \hline
                \end{tabular}
                \label{tab:compareNB}
\end{table}

The extraordinarily rapid scaling of the discovery index $N$ with the LSP mass, {\it cf.} Refs.~\cite{Li:2011gh,Li:2011rp}, which is basically
collinear with the primary input parameter $M_{1/2}$, highlights the comparative stability of the relative advantage garnered under the
ULTRA:9 selection cuts, as compared to the more standard CMS:3 variety.  Moreover, the rather small absolute values of $N$ which exist for the
ULTRA:9 selection cuts up to a level of about $m_{\rm LSP} \simeq 110$~GeV, {\it i.e.}~the vicinity of our leading benchmark $\cal{F}$-N, strikingly
exemplifies our claim that the vast majority of the No-Scale \fsu5 parameter space is already fully testable under the existing accumulation of data.
The most striking result of this tabulation, however, is the dramatic enhancement of the relative visibility, which jumps from order ten
in Table~(\ref{tab:compareN}), to up to order of one hundred in the present context.  This assumes, again, a sufficient modeling of the ultra-high
jet multiplicity SM backgrounds by the $t \overline{t} + {\rm jets}$ processes, a topic which will itself be revisited in Section~(\ref{sct:bgs}) following.

The results of Table~(\ref{tab:compareNB}) are translated into graphical form in Figure Set~(\ref{fig:Ultra_Ratio}), with the absolute discovery
indices plotted on the left, and the per-model comparative discovery ratio plotted on the right.
The disconnect between the continuity embodied in the No-Scale \fsu5 model space and the island CMSSM model SPS3 highlights the difficulty of
a head to head comparison between models with fundamentally different origins and spectra.  Of course, the fact that SPS3 fares ``better''
in terms of discoverability at a corresponding LSP mass is consistent with the fact that this model has already been ruled out.
Because of the distinctively light gluino and stop squark in No-Scale \fsu5, much lighter than all other squarks, these models tend to be
much more resilient against light squark limits than CMSSM models with comparably light LSP particles, and comparably strong SUSY production
cross sections.

Curiously, the plot of the comparative discovery ratio demonstrates that the relative advantage of the ULTRA:9
cutting philosophy is in fact maximized, reaching an extrapolated
peak of about $100$ times, in the immediate vicinity of model elements which were favored in
our Section~(\ref{sct:1p1ifb}) analysis of the most recent LHC data.  This region will be taken to constitute
a newly updated ``Golden Strip'' ({\it cf.} Ref~\cite{Li:2010mi}) of the bare minimally constrained~\cite{Li:2011xu} model space.
The region of the Golden Strip features an exceedingly satisfactory phenomenological agreement with limits on the flavor changing neutral current
$(b \rightarrow s\gamma)$ process, using a two standard deviation lower bound of $2.86 \times 10^{-4}$ on $Br(b \rightarrow s\gamma)$~\cite{Barberio:2007cr,Misiak:2006zs},
and likewise with limits on the anomalous magnetic moment of the muon, using a lower bound on the post-SM contribution $\Delta a_\mu$
to $(g-2)_\mu \div 2$ of $11 \times 10^{-10}$~\cite{Bennett:2004pv}, or more conservatively, of $9 \times 10^{-10}$.
Although both considered effects are at their lower limits at the strip boundary, they exert pressure in opposing directions on
$m_{\rm LSP} (or $$M_{1/2}$) due to the fact that the leading SUSY contributions to $Br(b\rightarrow s\gamma)$ enter with an opposing sign
to the SM term (requiring a sufficiently large mass that they not undo the SM component), while for the non-SM contribution to $\Delta a_\mu$,
the effect is additive (requiring a sufficiently small mass to make an appreciable contribution).

The complete \fsu5 model space is further easily consistent with the process $B_{s}^{0} \rightarrow \mu^+ \mu^-$, using an upper bound on
the branching ratio of $1.9 \times 10^{-8}$~\cite{Chatrchyan:2011kr}.  Collider based studies~\cite{Akeroyd:2011kd} of this process
are expected to continue to compete well with direct detection searches for this process, and place rather stringent limits on
certain sectors of the CMSSM.  However, these limits ease with decreasing $\tan \beta$, with an extremely strong
dependence in the sixth power.  No-Scale \fsu5 stably predicts a comparatively small ratio for $\tan \beta$,
in the vicinity of the value 20, which should not be impinged upon by any near term studies.  Likewise, our rather
heavy CP-Odd Higgs A, with $M_A$ in the range of $0.6 - 1.6$~TeV, contributes to a persistent immunity against this metric.
We emphasize that the heavy squarks of these models, having a mass of about 1~TeV, preserve the defining intent of SUSY
with respect to naturalness in stabilization of the gauge hierarchy.

The LSP within the full No-Scale \fsu5 model space is also quite satisfactory with regards
to the relevant spin-independent scattering cross-section bounds on Weakly Interacting
Massive Particles (WIMPs) from direct detection probes from XENON100~\cite{Aprile:2011hi}, and spin-dependent scattering cross-section limits from Super
Kamiokande~\cite{Tanaka:2011uf}, escaping the spin-dependent limits by three to four orders of magnitude.
This is facilitated, for the latter case in particular, by the fact that our models are
in the stau-LSP neutralino coannihilation region, with relatively heavy squark content.
The status of the Higgs boson search is also currently getting quite interesting,
with Summer conference reports from the ATLAS collaboration indicating
potential signals at about the $2.8~\sigma$ level peaking around
128~GeV and 144~GeV, and from the CMS collaboration indicating potential
signals at about the $2.0~\sigma$ level peaking around 120~GeV and 140~GeV.
The light Higgs in a large region of the \fsu5 model space is predicted to
have a mass of 120-128~GeV.  We note also here that the vector-like
particle mass $M_{\rm V}$, which may take a value up to several TeV,
and which does not directly couple to the Higgs, should make a comparatively
minor contribution to the Higgs mass.  A potentially relevant study of the general correlation
between the Higgs and a vector-like field multiplet has recently been released~\cite{LMNW-P}\cite{Evans:2011uq}.

\section{Elaboration on Standard Model Backgrounds\label{sct:bgs}}

Considering the large number of hadronic jets which are required by our optimized ultra-high jet signatures,
we have argued~\cite{Maxin:2011hy,Li:2011hr} that there is little intrusion from SM background processes after post-processing cuts.
Specifically, we have examined the background processes studied in~\cite{Baer:2010tk,Altunkaynak:2010we} and assessed the relevance
of each to our model in the initial LHC run. Our conclusion to date has been that only the $t \overline{t} + {\rm jets}$ process possesses
the requisite minimum cross-section and multiplicity of final state jet production to compete with the ${\cal F}$-$SU(5)$ signal.
Processes with a larger number of top quarks can also generate events with a large number of jets, however, the cross-sections have been deemed 
sufficiently suppressed to be negligible, bearing in mind the large number of ultra-high jet events which our model will generate.
Similarly, we have neglected in our Monte Carlo the pure QCD $(2,3,4)$ jet events, one or more vector boson events, and all $b \overline{b}$ processes,
since none of these have been judged capable of sufficient event production with 9 or more jets after post-processing cuts have been applied.
The same has been taken to hold for those more complicated background processes involving combinations of top quarks, jets, and one or more vector bosons, which yield a very large number of raw events, though it has been expected that practically all of the jets from detector effects beyond the
initial hadronization would be ultimately discarded.  We revisit these conclusions in the present section, and in particular, consider
the expected consequences for discoverability of our model if these assumptions should be in various regards materially incorrect or incomplete.

It is not immediately clear how the published CMS results for $\ge$9 jets with $p_T \ge$ 50 GeV translates into the context of a reduced threshold on the transverse momentum $p_{\rm T}$, and we shall continue to study the issue.  It seems for now though that we can set limits based on a set of worst case assumptions, comparing the CMS results with our own $t \overline{t} + {\rm jets}$ simulation.  Following this lead, if we scale the QCD multijet background at three times the combined $t \overline{t},W^\pm,Z + {\rm jets}$ contribution in Figure~(\ref{fig:N_Jets}), of which nearly all is taken to be modeled by $t \overline{t} + {\rm jets}$ in the Ultra-High jet regime, the net result is a 4 times upgrade of the background.  This reduces the average discovery advantage $N {\rm (ULTRA:9)} / N {\rm (CMS:3)}$ from the value of 44.1 given in Table~(\ref{tab:compareNB}) to a value of 15.4; perhaps not so surprisingly, this is essentially the titular result of order ten from Table~(\ref{tab:compareNB}), based on a na\"{\i}ve global application of just the $t \overline{t} + {\rm jets}$ sampling.  If we assume the combined $t \overline{t},W^\pm,Z + {\rm jets}$ contribution is only 25\% modeled by the $t \overline{t} + {\rm jets}$, then a multiplicative factor of 16.0 is applied to the backgrounds,
and the ULTRA:9 discovery advantage is still a healthy 4.1.

\section{Conclusion and Summary}

We seem now to be firmly entering the golden age of LHC physics.  The collider is working brilliantly,
and exceeding scheduled goals for the ramping up of integrated luminosity. By recently crossing the
threshold of one delivered femtobarn of integrated luminosity, this remarkable machine begins to dramatically
reshape our perceptions of the plausible landscape of Supersymmetric extensions to the Standard Model,
upgrading the initial reports based upon only $35~{\rm pb}^{-1}$ of integrated luminosity by more than 30 times.
The present work marks our first opportunity to comment on the ongoing LHC search following the first presented
analysis~\cite{PAS-SUS-11-003} of data eclipsing the $1.1~{\rm fb}^{-1}$ milestone, but likewise extends the scope of
prior work in several other regards.  In particular, we have combined an established
Monte Carlo simulation of the full bare minimal parameter space of the No-Scale \fsu5 model class, at five distinct collider
energies with the recently introduced discovery index statistic $N$, for four specific cut methodologies, two of which are
considered for the first time.  In addition, we supplement our limited background
simulation with actual collider results harvested from the most recent LHC observations.

Several new key results have emerged during the current study.  We have established the first exclusion boundaries on
the bare-minimally constrained model space of No-Scale \fsu5, resulting from the first 1.1 inverse femtobarns of integrated
LHC luminosity.  We find that the LSP mass in these models should
be at least about $92$~GeV, with a corresponding boundary gaugino mass $M_{1/2}$ above about $485$~GeV, as characterized
by the benchmark $\cal{F}$-J.  We find the optimal fit to occur at somewhat heavier models, including a very suitable
benchmark located at $m_{\rm LSP} = 108.8$~GeV and $M_{1/2} = 560$~GeV, named $\cal{F}$-N.  Furthermore, in contrast to higher mass constraints in the CMSSM, we found lower limits on the gluino and heavy squark masses in the \fsu5 model space in the range of 658-674 GeV and 854-1088 GeV, respectively, with the minimum boundary on the light stop mass at about 520 GeV. Not only are the models in
the vicinity of these points capable of adroitly escaping the onslaught of LHC data which is currently decimating the standard
mSUGRA/CMSSM benchmarks, they are also able to efficiently explain certain tantalizing production excesses over the
SM background which have been reported by the CMS collaboration.  Critically also, a clear path is provided to salvage the
defining motivation of Supersymmetry itself, that being a natural stabilization of the gauge hierarchy, as embodied in sparticle
mass splittings of not more than about 1~TeV.

We have emphasized throughout the simple but rather critical observation that different model classifications
respond differently to various alternative selection cut criteria.  In particular, results which legitimately discount
substantial segments of the Minimally Constrained Supersymmetric Standard Model must not be inferred to also do similar
damage to the underlying framework of Supersymmetry itself.  This is because: 1) the CMSSM represents a simplification which
is sometimes necessary for the convenience of the analyst, but perhaps not so for nature herself; 2) it is simply quite difficult
to make fair comparisons between contending models -- one might say after all that this is why they are in fact called different
models; and 3) again, critically, any subtle reordering of the model spectra may translate to substantially
differential signal responses to the chosen selection cuts.  Highlighting these observations, we have here expanded our study of
a proposed set of selection cuts designed to reveal the natural ultra-high jet multiplicity signal associated with the stable mass
hierarchy $m_{\tilde{t}} < m_{\tilde{g}} < m_{\tilde{q}}$ of the \fsu5 models.  It has been demonstrated that an enhancement of
order ten in model visibility may be attained by adoption of these cuts, which is remarkably stable in simulation across the
\fsu5 model space, and likewise also for a sampling of various upgraded LHC beam energies.  This factor is sufficient to
immediately and definitively test a majority of the No-Scale \fsu5 model space, using only the already collected LHC data set.

We have stressed that habits established in lower jet multiplicity searches with regards to the appropriate kinematic thresholds per jet and missing 
energy diagnostics such as $\alpha_{\rm T}$ and $\Delta \phi^*$ do not necessarily translate well into the ultra-high
jet multiplicity search regime.  The extraordinary cost, both of labor and capital, exerted thus far in the LHC effort argue vigorously that every
available efficiency which may be freely rendered from updates in the methods of analysis should be promptly seized up.
As the long-favored oases of the CMSSM framework evaporate before our eyes, the time approaches rapidly when alternative
search criteria must be implemented, designed to uniquely illuminate new dreams within the broader SUSY philosophy.
The extraordinary potential discoverability of No-Scale \fsu5, apparent though only under application of the appropriate selection
tool, is at the heart of our claim to a model representing the dichotomy of imminent testability combined with a remarkable
resilience of viability in the face of all existing testing.


\begin{acknowledgments}
This research was supported in part 
by the DOE grant DE-FG03-95-Er-40917 (TL and DVN),
by the Natural Science Foundation of China 
under grant numbers 10821504 and 11075194 (TL),
by the Mitchell-Heep Chair in High Energy Physics (JAM),
and by the Sam Houston State University
2011 Enhancement Research Grant program (JWW).
We also thank Sam Houston State University
for providing high performance computing resources.
\end{acknowledgments}


\bibliography{bibliography}

\end{document}